\documentclass[twocolumn]{aastex631}
\usepackage{graphicx}
\graphicspath{{images/}}
\usepackage{enumerate}
\usepackage{amssymb, amsmath, amsfonts}
\usepackage{mathtools}
\usepackage{natbib}
\usepackage{color}
\usepackage{hyperref}
\usepackage{ulem}
\usepackage{soul}
\usepackage{booktabs}
\usepackage{multirow}
\usepackage{graphicx}
\usepackage{enumitem} 
\setitemize{noitemsep,topsep=0pt,parsep=0pt,partopsep=0pt}

\newcommand{\tglc}{{\tt tglc}}

\def\arraystretch{1.1}

\begin{document}

\title{TESS-Gaia Light Curve: a PSF-based TESS FFI light curve product}

\author[0000-0002-7127-7643]{Te Han} \affiliation{University of California, Irvine}
\affiliation{University of California, Santa Barbara}

\author[0000-0003-2630-8073]{Timothy D.~Brandt} \affiliation{University of California, Santa Barbara}

\begin{abstract}
The Transiting Exoplanet Survey Satellite (TESS) is continuing its second extended mission after 55 sectors of observations. TESS publishes full-frame images (FFI) at a cadence of 1800, 600, or 200 seconds, {allowing light curves to be extracted} for stars beyond a limited number of pre-selected stars. Simulations show that thousands of exoplanets, eclipsing binaries, variable stars, and other astrophysical transients can be found in these FFI light curves. To obtain high-precision light curves, we forward model the FFI with the effective point spread function to remove contamination from nearby stars. We adopt star positions and magnitudes from Gaia DR3 as priors. The resulting light curves, called TESS-Gaia Light Curves (TGLC), show {a photometric precision closely tracking the pre-launch prediction of the noise level. TGLC's photometric precision reaches $\lesssim$2\% at 16th TESS magnitude even in crowded fields.} We publish {TGLC Aperture and PSF light curves} for stars down to 16th {TESS} magnitude through the Mikulski Archive for Space Telescopes (MAST) for all available sectors and will continue to deliver future light curves via \dataset[10.17909/610m-9474]{\doi{10.17909/610m-9474}}. The open-source package $\tglc$ is publicly available to enable any user to produce customized light curves. 

\keywords{Light curves (918) --- Astronomy software (1855) --- Astronomy databases (83) --- Exoplanet astronomy (486) --- Variable stars (1761) --- Eclipsing binary stars (444)}
\end{abstract}

\section{Introduction} \label{sec:intro}
The Transiting Exoplanet Survey Satellite (TESS) offers {nearly complete} sky coverage, uniform cadence, and high-precision photometry.  This enables a huge amount of time-domain science, from transiting planets \citep{piMensae,Vanderspek_2019}, to eclipsing binaries \citep{TESS_eclipsing_binaries,sextuple_eclipse,tripleeclipse}, to stellar pulsations and variability \citep{deltaScuti,WDpulsations,asteroseismology}, to exotic binaries with accretion and complex variability \citep{CVs,TESS_AMCVn,postcommonenvelope}, to blazars \citep{blazars_old,blazars} and {supernova} transients \citep{supernovae_1,supernovae}. The headline TESS science product is its two-minute photometry {processed with the official Science Processing Operations Center (SPOC) pipeline \citep{SPOC}}, but this is only available {for $\sim$10$^5$ stars} in each 27-day observed sector. Most of the TESS data volume consists of full-frame images {(FFIs)} binned to an 1800-second, 600-second, or 200-second cadence. {Many of the aforementioned science cases require extraction of light curves from the FFIs for stars other than the pre-selected targets.} 

TESS has the capability of reaching a photometric precision of $\approx$10$^{-2}$ in a 30-minute exposure at 16th {TESS} magnitude \citep{TESS}. This matches the per-epoch performance of the Zwicky Transient Facility \citep[ZTF,][]{ZTF}, but TESS offers $\approx$1000 measurements over $\approx$27 days of nearly continuous viewing. However, a user seeking a TESS light curve for a 16th {TESS} magnitude star must currently download and process the raw full-frame images or relevant subimages with the help of TESScut \citep{TESScut} and, potentially, with a package like {\tt eleanor} \citep{eleanor}. \cite{eleanor_lite} published all TESS FFI {corrected aperture} light curves for stars brighter than 16th {TESS} magnitude without flux contamination removal, which is essential to produce reliable light curves for dim stars.

Several pipelines have published individual stars' light curves on the Mikulski Archive for Space Telescopes (MAST). {However, the {Quick-Look Pipeline} \citep{QLP} and the {SPOC} full-frame images pipeline \citep{SPOC_FFI} only provide light curves for stars brighter than 13.5 {TESS} magnitude; the Cluster Difference Imaging Photometric Survey (CDIPS) and the PSF-based Approach to TESS High quality data Of Stellar clusters (PATHOS) only provide a subset of TESS data \citep{CDIPS,PATHOS}; {\tt eleanor} and the TESS Data for Asteroseismology collaboration can return a light curve of any star in a full-frame image at the price of a degree of user involvement.} Many of these pipelines remove contamination from nearby stars and adjust for the background, but systematics do remain. For all of these pipelines, the large size of TESS pixels ($\approx$21$\arcsec$) is an intrinsic limitation that is especially problematic for faint stars and crowded fields. 

In addition to the large size of TESS pixels and consequent blending from nearby stars, TESS also has fluctuating backgrounds from scattered sunlight, especially during certain phases of its orbit. {Users interested in precise, high-cadence photometry for a fainter star must overcome all of these limitations by post-processing the full-frame images.} The best extracted light curves for stars fainter than 13.5 {TESS} magnitude continue to require significant user input \citep{eleanor}. 

This paper presents a method to produce uniform, calibrated, light curves for all stars brighter than 16th {TESS} magnitude by leveraging astrometry and photometry from the Gaia mission. TESS and Gaia have full or nearly full-sky coverage, but their capabilities are complementary. Gaia has measured precise positions and brightnesses of {about 1.5 billion stars} with a Gaia Rp filter similar to that of TESS \citep{Gaia_EDR3}. Gaia's angular resolution of $\approx$0$.\!\!\arcsec2$ is a factor of 100 higher than TESS's, while TESS brings a high, uniform cadence with nearly continuous coverage over its observations of each 27-day sector. The precise measurements of Gaia can enable TESS to overcome its limitations of poor angular resolution and high, fluctuating backgrounds. We use Gaia astrometry and photometry to constrain the field stars in a TESS image, build a complete and local {point spread function forward model} of each TESS full-frame image, and extract percent-level precise light curves of {approximately 3 million stars per sector} down to 16th {TESS} magnitude. {The final light curves of all sectors are being published in MAST's High Level Science Product (HLSP) database continuously with the release of TESS FFI.} We name them TESS-Gaia Light Curves (TGLC).

The paper is structured as follows. Section \ref{sec:existing} discusses the existing TESS FFI pipelines and their limitations. Section \ref{sec:method} explains our pipeline's methodology. Section \ref{sec:lc_extraction} examines two types of TGLC light curves. We analyze the photometric precision of TGLC in Section \ref{sec:comparisons}. Section \ref{sec:exoplanet} presents a case study of five known exoplanets using TGLC. Section \ref{sec:data} describes the publication of our data product and python package.  Finally, we discuss the influence of TGLC on TESS time-domain science in Section \ref{sec:discussion}. 

\section{Existing TESS FFI light curve products} \label{sec:existing}
\begin{figure*}
    \centering
    \includegraphics[width=\textwidth]{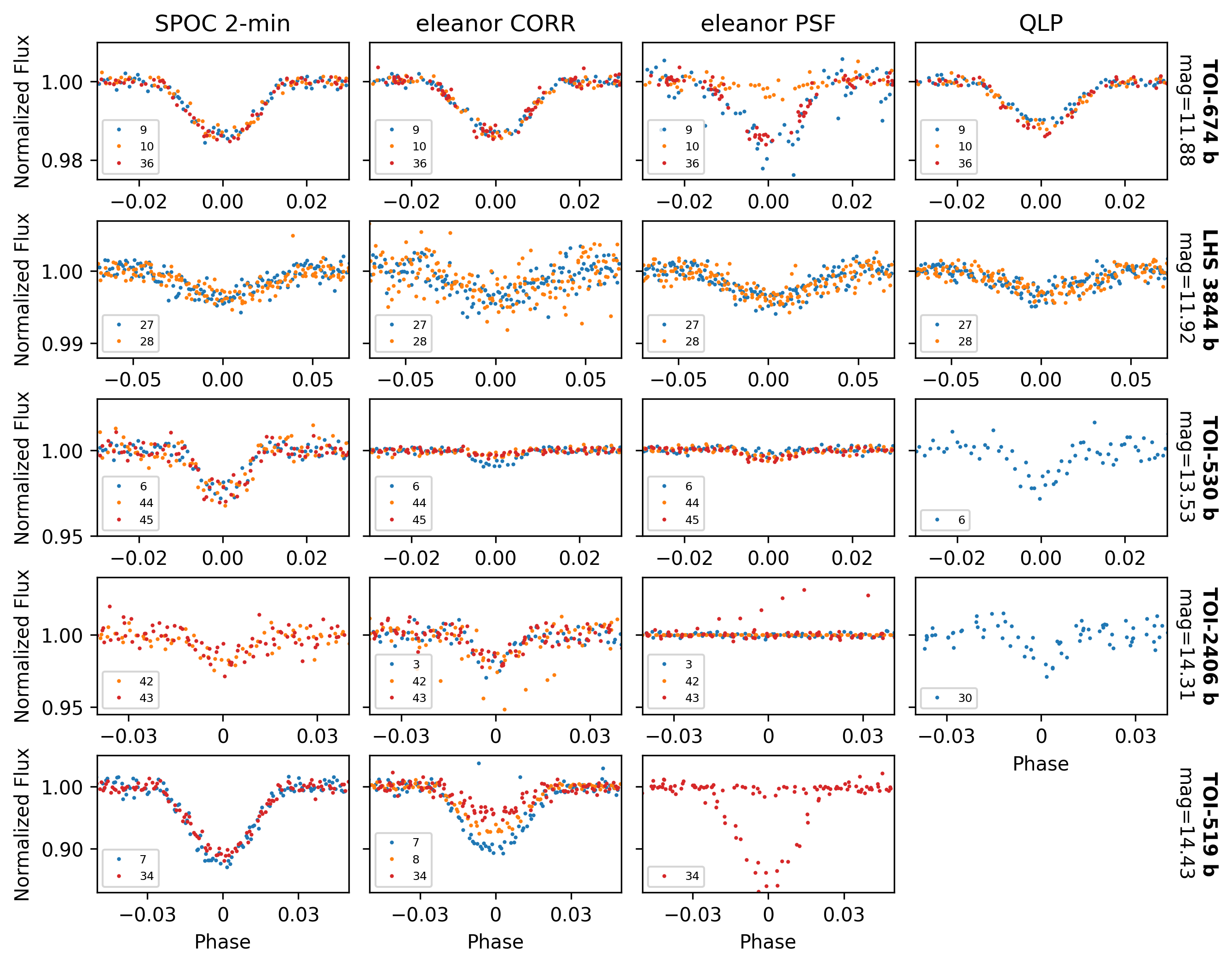}
    \caption{Light curves of five TESS-discovered exoplanets from existing pipelines. Compared to the published work (similar to the SPOC 2-min light curves), the current FFI light curves perform poorer, especially for dimmer stars. The legend in each subplot indicates the sector(s) plotted. {Light curves from extended missions are binned to a 30-minute cadence for compatible noise level with the primary mission. The periods and transit midpoints are adopted from each discovery literature.} All lightcurves are detrended with \textsf{wotan} \citep{wotan} and the phases are normalized to 1. }
   \label{fig:exoplanet_other}
\end{figure*}

In this section, we give an overview of the existing sources of FFI light curves, focusing on two that provide magnitude-limited samples: the Quick-Look Pipeline \citep{QLP}, and {\tt eleanor} \citep{eleanor}. The Quick-Look Pipeline (QLP) light curves are produced with an aperture approach combined with difference imaging.  A template is first constructed by combining many comparison frames; each FFI is then differenced relative to this comparison frame.  QLP performs aperture photometry on this difference image and then uses the TESS magnitude to scale the difference relative to the star's average flux. However, the nature of aperture photometry involves a tradeoff: a larger aperture captures more of a star's photons, but at the cost of increased backgrounds, {contaminations,} and, potentially, systematics.  The performance of QLP also relies on the fidelity of its template and, as a result, on the temporal stability of systematics.

Another popular TESS FFI light curve product, {\tt eleanor}, {has two versions of calibrated light curves:} principal component analysis (PCA) {and} point spread function (PSF) light curves \citep{eleanor}. {The aperture photometry} {\tt eleanor} PCA utilizes the co-trending basis vectors published by the Science Payload Operations Center (SPOC, \cite{10.1117/12.2233418}) to remove systematics in each camera. {{\tt eleanor} PCA light curves are further calibrated to produce {\tt eleanor} CORR.} {\tt eleanor} PSF uses an analytical two-dimensional Gaussian model to perform PSF photometry. Users can incorporate positions of several nearby stars in the PSF fit to remove contamination partially, but inputting each star's position and magnitude in the fit manually becomes impractical for a full-sky light curve product. The {\tt eleanor} documentation also suggests summing multiple two-dimensional Gaussians to accommodate irregular PSF shapes. However, the results are still not ideal for some irregular PSF shapes and require a higher level of user involvement. 

Figure \ref{fig:exoplanet_other} compares the light curve products of three pipelines: {\tt eleanor} CORR, {\tt eleanor} PSF, and QLP, to the {official} SPOC 2-min light curves.  We adopt five faint TESS planet hosts as our comparison stars: TOI-674 \citep{Murgas_2021}, LHS 3844 \citep{Vanderspek_2019}, TOI-530 \citep{Gan_2021}, TOI-2406 \citep{Wells_2021}, and TOI-519 \citep{Parviainen_2020}. We plot the light curves phase-folded around the center of their transits. The SPOC light curves are compared because all five publications above use the SPOC pipeline in their discoveries of the new exoplanets. All light curves {from extended missions} are binned to {a} 30-min cadence {like the primary mission} for a fair comparison of noise levels. {All light curves are detrended with \textsf{wotan}, an automated detrending algorithm \citep{wotan}. It is especially powerful in preserving transit signals while removing stellar trends. The detrending method is biweight, and the window length is set to 1 for all detrended light curves in this paper.} For the first two stars of $\approx$12 {TESS} magnitude, the FFI light curves mostly perform comparably to the 2-min light curves, but they become visibly worse for the dimmer stars. The light curves for these fainter stars have either a lower SNR, an inaccurate transit depth (compared to published work), inconsistency among sectors, or a combination of these issues. 

\section{Point Spread Function Modeling of TESS Full-Frame Images} \label{sec:method}
Point spread function (PSF) photometry is the most accurate way to obtain brightnesses, provided a sufficiently good PSF model is available. The PSF varies spatially across TESS fields, so PSF models must be fit locally to address the variation. PSFs can also be time-dependent, so they need to be fit for each frame. However, building an accurate PSF model locally and frame-wise for TESS is daunting. Without prior knowledge of the positions and brightnesses of stars in the field, PSF fitting is hopelessly underconstrained: the number of free parameters of star positions and magnitudes can exceed the number of pixels {because of the large TESS pixels}. {We have $\sim$ 1.5 billion stars in Gaia DR3, and each pixel is approximately $10^{-8}$ sr, which implies $\sim$1.2 stars per pixel or $\sim$3.6 parameters to constrain with one pixel value on average.} Any fit for the PSF and the stellar positions would also be fundamentally nonlinear. Even if one were to ignore enough low-brightness stars to reduce the number of free parameters, the computational performance of such a nonlinear fit would still be impractical for a full-sky survey.

Gaia’s astrometry and photometry offer a way to overcome this inability to measure the instrumental PSF. If we apply precise Gaia position and brightness measurements as fixed priors, we are left with only the PSF parameters. We can construct this problem as a linear fit, solvable with standard linear algebra-based approaches as shown below.  Furthermore, we can include all Gaia stars in the forward modeling to resolve the background {contamination as deep as $\sim$20 {TESS} magnitude} without unduly compromising computational performance. 

The fundamental assumption of our PSF modeling is that the PSF shape is spatially constant over each FFI cutout of the size of {about $150 \times 150$ pixels} for each epoch. In practice, the PSF variation is gradual among adjacent cutouts as shown in Section \ref{sec:fitting_models}, and applying the same PSF within one cutout is reasonable. We also allow the PSF to vary with time by fitting each frame independently. We first extract stellar positions and brightnesses from Gaia DR3. {We then fit the PSF parameters of the simulated image to the FFI cutout.} Using the best-fit PSF shape, we forward-model the FFI cutout with all stars except the target. Lastly, we generate a decontaminated image by subtracting this modeled cutout from the measured FFI cutout. We can then perform aperture photometry or PSF photometry on the residual to produce photometry of a given star at a given epoch.  By repeating this process for all stars and all frames, we can construct FFI light curves for all stars in TESS.

\subsection{Positions and Brightness from Gaia}

The Gaia mission has now measured positions and magnitudes of {about 1.5} billion stars \citep{Gaia_EDR3}. The positions are typically accurate to $\lesssim$0.1 mas or better, while magnitude uncertainties remain $\sim$0.001 mag even down to a Gaia {G} magnitude of 18. This extraordinary data set enables us to construct a full forward model of each TESS full-frame image. 

A forward model requires the positions and TESS magnitudes of all stars. {We first propagate Gaia positions from J2016.0 to the median TESS epoch of each sector using the measured Gaia proper motions. We then convert the Gaia DR3 right ascension and declination to pixel positions based on the TESS FFI world coordinate system (WCS) headers.} We do not account for either parallactic motion or perspective acceleration {because} both are typically orders of magnitude smaller than the proper motion correction and $<0.01$ TESS pixels for all but a handful of stars within a few parsecs. The positions could have small, local offsets from the true coordinates {due to the TESS spacecraft jitter motion}. As we will show later, however, photometric products from our PSF approach are largely immune from systematic shifts over FFI cutouts.  We rely only on the accurate relative positions in the TESS frame {across} $\approx$150 TESS pixels. 

In addition to astrometry, our approach also requires photometric anchors for all stars in the field.  The TESS bands differ from the Gaia bands, requiring a color-dependent transformation.  We adopt the conversion
\begin{align} \label{eq1}
    T = &\; G - 0.00522555(G_{\text{BP}} - G_{\text{RP}})^3 \nonumber \\ &+ 0.0891337(G_{\text{BP}} - G_{\text{RP}})^2 \nonumber \\ &- 0.633923(G_{\text{BP}} - G_{\text{RP}}) + 0.0324473
\end{align}
published in \cite{Stassun_2019}, where $T$ is the apparent magnitude in the TESS bandpass, and $G$, $G_{\text{BP}}$, and $G_{\text{RP}}$ are Gaia bandpasses. When $G_{\text{BP}}$ and $G_{\text{RP}}$ are missing, we adopt
\begin{equation} \label{eq2}
    T = G - 0.430
\end{equation}
from the same {work}. After this step, we have astrometry and photometry in the TESS system for all Gaia stars.

\subsection{Effective PSF Model}
\begin{figure}[t]
    \centering
    \includegraphics[width=\linewidth]{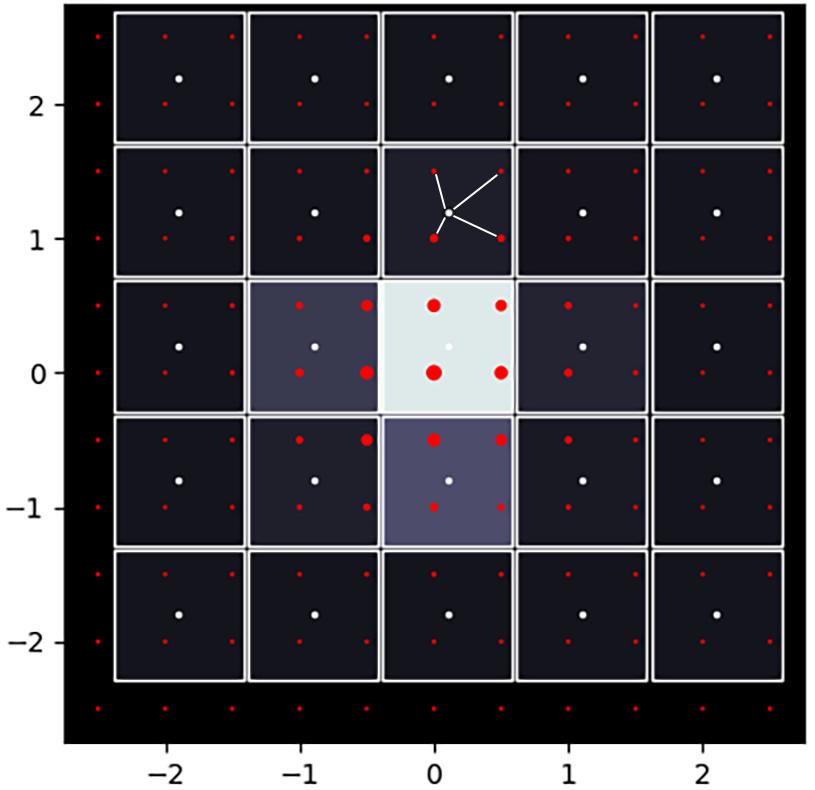}
    \caption{Interpolation example of an ePSF model. The white squares and dots represent the pixels and their centers; the red dots represent the twice oversampled ePSF model; the dot sizes represent the ePSF values. The star represented by the red ePSF values is located at $(0, 0)$, the location of the largest red dot. Each pixel value (grayscale) is interpolated only by its nearest four values in the oversampled ePSF (red points) as shown in pixel (0, 1) by the white lines. This maintains the linearity (and therefore the computational efficiency) of the model we use to forward model a subimage of a full-frame TESS image from an ePSF. }
   \label{fig:ePSF}
\end{figure}

With positions and magnitudes known for all stars in a TESS field, we can forward model a full-frame image. We approach the problem agnostic to the form of the PSF, using the effective PSF (ePSF) model successfully applied to Hubble data \citep{Anderson+King_2000}. The ePSF model notes that the actual PSF observed on the detector is the convolution of the PSF incident on the detector and the detector response function. This ePSF is continuous and is the only observable form of the PSF. The ePSF may be defined at anchor points spaced more finely than the pixels on the detector and then interpolated at the position of a source. Interpolating the ePSF is equivalent to placing the PSF produced by the optics at the location of a star, multiplying by the pixel response function, and then integrating over the pixels \citep{Anderson+King_2000}.

Figure \ref{fig:ePSF} shows how the ePSF model works. The ePSF is shown at red anchor points and {the simulated pixel values are interpolated among those red points}. In the example figure, the ePSF is oversampled relative to the pixels on the detector by a factor of two. The size of the red points gives the intensity of the ePSF at that point, with the star centered at the point $(0,0)$. The data are interpolated from the ePSF at the pixel centers, indicated in white. The grayscale color over the pixels shows these interpolated values from the ePSF, which indicate the actual intensity observed for a point source centered at $(0, 0)$, which in this case is displaced from the center of a pixel.

The effective PSF (ePSF) method from \cite{Anderson+King_2000} interpolates the ePSF from a four-times-oversampled grid with cubic spline interpolation. Given the arbitrary nature of the oversampling factor, we explore oversampling factors from 2, 4, and 6. We decide to use a factor of 2 because it produces the best quality light curves (those derived in Section \ref{sec:lc_extraction}) and is much more computationally efficient. 

We also use bilinear instead of cubic interpolation.  This choice enables us to write the optimization of the ePSF's values at its anchor points as a linear least-squares problem.  This fact is necessary to keep the model computationally tractable. Also, we assume that the PSF of a star is mostly contained within a square with a side length of 11 pixels. Fluxes in further pixels are negligible according to TESS Instrument Handbook v0.1 \footnote{\url{https://archive.stsci.edu/files/live/sites/mast/files/home/missions-and-data/active-missions/tess/_documents/TESS_Instrument_Handbook_v0.1.pdf}}. In a grid with side length $L_{\text{grid}} = 11$, the oversampled grid side length is $rL_{\text{grid}} + 1$, where $r$ is the oversampling factor. Depending on the position of a star to the center of its pixel, stars are interpolated by different sets of four interpolation points. We can now model all stars in a cutout by fitting a shared local ePSF while fixing stars' positions and relative fluxes, but we must also consider the background to model the field completely. {The derivation of our complete model is presented in Section \ref{sec:fitting_models} after the discussion of background modeling. }

\subsection{Background modeling} \label{sec:background}
\begin{figure*}
    \centering \includegraphics[width=\textwidth]{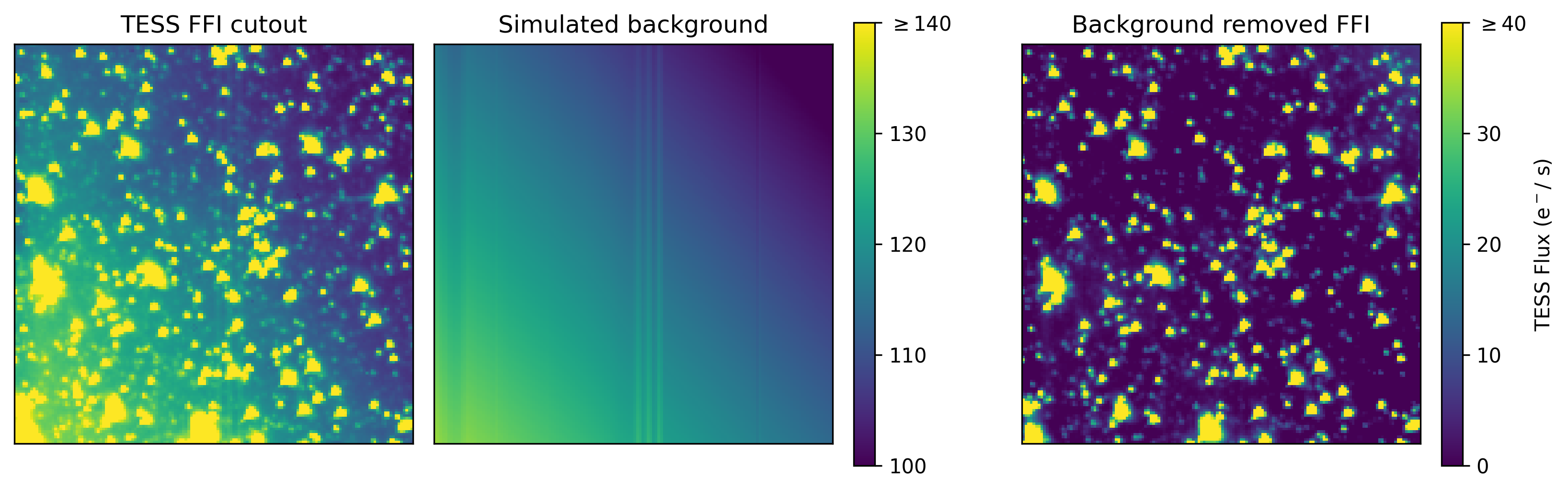}
    \caption{Example of background removal of TGLC. The left panel shows a FFI cutout with a background gradient and three vertical straps. We model the background as the middle panel. The residual image on the right shows a much cleaner field ready for an ePSF fit. Note that we preserve the resolution of the color map and only shift it down by 100 $e^-/s$ for the last panel.}
  \label{fig:cal_bg}
\end{figure*}

The unevenness of the TESS FFI background is a result of both stray light and CCD artifacts. Stray light from the Earth and the Moon usually produce a background with a gradient and are strongest near the ends of each observation window. The gradient is mostly linear in a small cutout and easy to fit. The CCD artifacts, including the straps shown in the first panel in Figure \ref{fig:cal_bg}, result from highly reflective metal straps beneath the CCD silicon. These straps, however, are strictly column-wise according to the TESS Instrument Handbook v0.1. We model the straps as a column-dependent flatfield that applies only to a portion of the background.  The visible effects of the straps are chromatic. They result from photons that penetrate the CCD, are reflected, and then detected; wavelengths where the CCD is more transparent show stronger artifacts at the strap locations according to TESS Instrument Handbook v0.1.

While the straps reside at specific CCD columns, they can reflect photons into neighboring columns.  We, therefore, model the effect of the straps as a column-dependent modification to the background.  We construct a flat, column-dependent background by fitting the variation in the background intensity between neighboring columns, applying a low-pass filter, and averaging across sectors for a given CCD. We take the dimmest half of the pixels in each column and compute the median of the ratio of the count rate between neighboring columns, using only pixels that are among the dimmest half in both columns.  Multiplying these ratios across the detector produces an effective column-dependent background.  We remove drifts in the count rate across the detector by dividing by a median-filtered processing of the background.  Finally, we take the median for each CCD's calibrated background over all primary mission sectors, since the straps are inherent to the CCD. 

The reflective straps on the back of the CCD have a strong chromatic effect.  To account for this, we use our calibrated CCD-dependent background as a flatfield for only a fraction of the background.  By allowing this fraction to be free, we fit the chromaticity of the strap reflections frame-by-frame.  We ultimately use six parameters to model the background: a flat background and a linear gradient in each of two dimensions, plus these same parameters multiplying the CCD-dependent calibrated background. Figure \ref{fig:cal_bg} shows an example model of the background: both the linear gradient and straps have been removed from the FFI. In particular, strap removal is essential for avoiding systematic vertical lines in ePSF shapes.

\subsection{Fitting ePSF and background models} \label{sec:fitting_models}
\begin{figure}
    \centering
    \includegraphics[width=\linewidth]{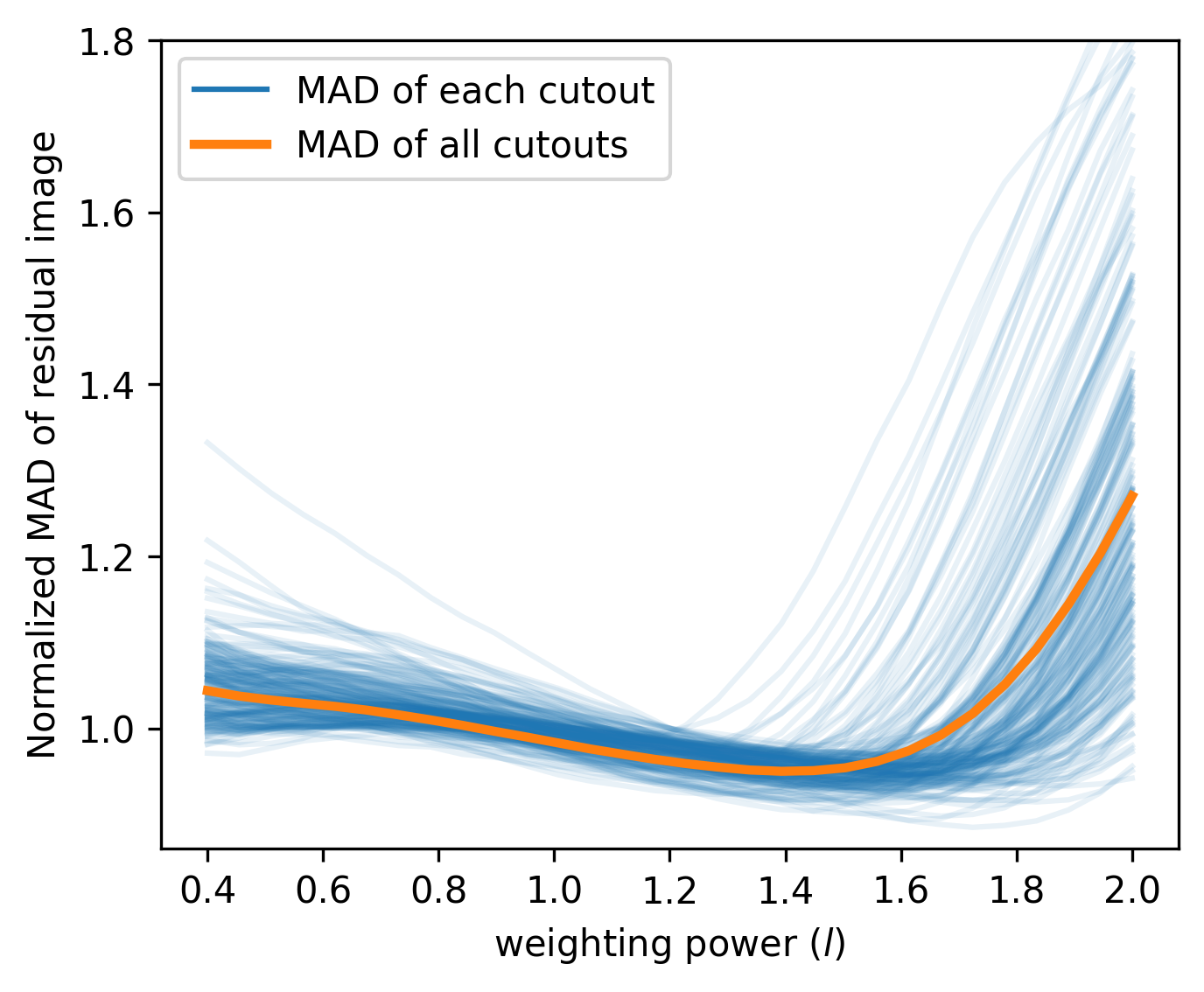}
    \caption{Normalized median absolute deviation (MAD) of residual images with different weighting power $l$ (Equation \eqref{eq:fullmodel_matrix_with_p}).  Larger values of $p$ weight lower count-rate pixels more heavily in deriving the ePSF, with $p=1$ weighting all pixels equally. These blue curves are generated from 196 different cutouts of Sector 7. The orange curve shows the MAD taken over all pixels from 196 residual images, which favors a weighting power $l \approx 1.4$}
   \label{fig:weight}
\end{figure}
\begin{figure*}
    \centering
    \includegraphics[width=\textwidth]{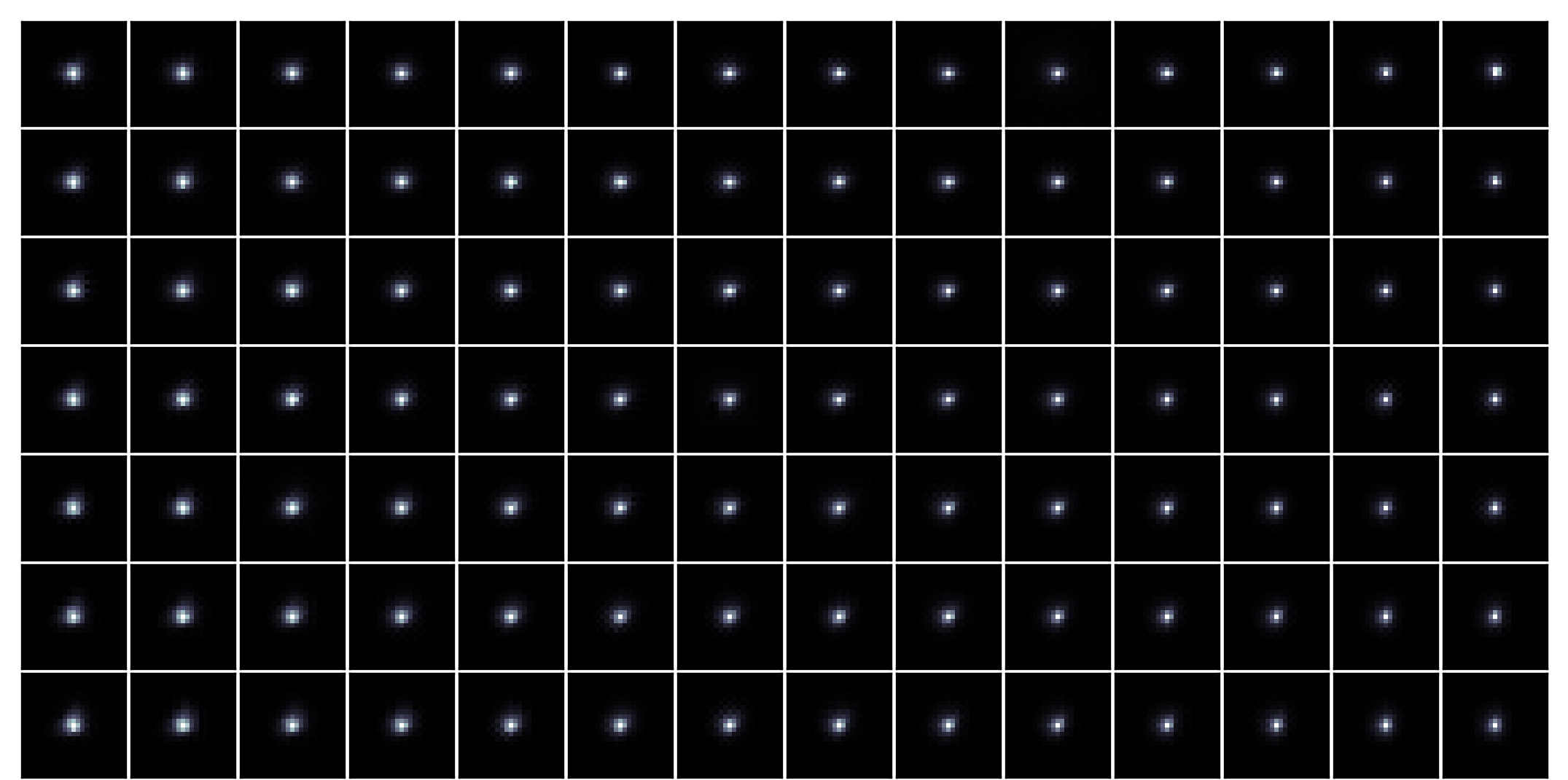}
    \caption{$7 \times 14$ effective PSF models for half of Sector 1, camera 4, CCD 3. Each ePSF model is fitted in a $150 \times 150$ pixels FFI cutout, and $14 \times 14$ models will exactly cover the $2048 \times 2048$ pixels image with two-pixel-wide overlaps. Each CCD is divided in this manner to produce light curves published on MAST. We can observe the gradual spatial variation of ePSF: An obvious trend is that the ePSFs are narrower in the upper right corner{, which is the closest to the center of the lens}. {One plausible reason is the optics of the telescope produce more compact pixel response functions (PRF) near boresights according to TESS Instrument Handbook v0.1.} The gradual variation supports our assumption of constant ePSF in each cutout at the beginning of Section \ref{sec:method}. }
   \label{fig:epsf_examples}
\end{figure*}

Obtaining the ePSF requires a simultaneous fit to all of the parameters describing the ePSF itself together with the parameters that model the background. We construct a least squares fit $P \approx AF$ as
\begin{equation}
  \left[ {\begin{array}{c}
    p_{1}\\
    p_{2}\\
    \vdots\\
    p_{m}\\
  \end{array} } \right] \approx
  \left[ {\begin{array}{cccc}
    a_{11} & a_{12} & \cdots & a_{1n}\\
    a_{21} & a_{22} & \cdots & a_{2n}\\
    \vdots & \vdots & \ddots & \vdots\\
    a_{m1} & a_{m2} & \cdots & a_{mn}\\
  \end{array} } \right]
  \left[ {\begin{array}{c}
    f_{1}\\
    f_{2}\\
    \vdots\\
    f_{n}\\
  \end{array} } \right],
  \label{eq:fullmodel_matrix}
\end{equation}
where $p_1$ to $p_m$ are the observed count rate of each pixel {from the FFI}, $a_{11}$ to $a_{mn}$ are the matrix encoded with star information, and $f_{1}$ to $f_{n}$ are best fits of both the ePSF and background parameters. {$m$ is the total number of pixels in the FFI cutout, and $n$ is the total number of free parameters. Each star only adds weights to the rows of $A$ (pixels) that are 5 pixels or closer to the star; the weights for each column are the bilinear translation weights from the oversampled ePSF grid to the corresponding pixel. Weights from all stars are summed to construct the full $A$. We solve for $F$ in the least square fit. } For example, if we fit for an $11 \times 11$ pixel ePSF (extending five pixels from the star) and oversample by a factor of two, this requires fitting $n = 535$ free parameters ($(11\times2+1)^2+6$). We need the number of pixels in a cutout to be much larger than the number of free parameters and for there to be many stars in the image for the fit to be over-constrained. Adopting $150 \times 150$ pixel regions ($m=22500$) provides more than enough pixels to constrain the ePSF, while such a region usually has at least thousands of stars detected in Gaia. 

Equation \eqref{eq:fullmodel_matrix} cannot be solved exactly if there are more pixels than parameters.  The best solution depends on the definition of ``$\approx$''.  To allow a linear algebra solution, we take it to mean the sum of the squares of the weighted differences between the observed and modeled pixel values.  In classic $\chi^2$ statistics, these differences are normalized by the observational uncertainty.  In our case, we explore different weightings to avoid heavily prioritizing the brightest pixels and to limit our vulnerability to detector nonlinearities. We parametrize a family of weights by an exponent $l$, weighting each residual by $1/p^l$.  This weighting modifies Equation \eqref{eq:fullmodel_matrix} to read
\begin{equation}
  \left[ {\begin{array}{c}
    p_{1} / p_{1}^{l}\\
    p_{2} / p_{2}^{l}\\
    \vdots\\
    p_{m} / p_{m}^{l}\\
  \end{array} } \right] \approx
  \left[ {\begin{array}{cccc}
    a_{11} / p_{1}^{l} & \cdots & a_{1n} / p_{1}^{l}\\
    a_{21} / p_{2}^{l} & \cdots & a_{2n} / p_{2}^{l}\\
    \vdots & \ddots & \vdots\\
    a_{m1} / p_{m}^{l} & \cdots & a_{mn} / p_{m}^{l}\\
  \end{array} } \right]
  \left[ {\begin{array}{c}
    f_{1}\\
    f_{2}\\
    \vdots\\
    f_{n}\\
  \end{array} } \right].
  \label{eq:fullmodel_matrix_with_p}
\end{equation}
where we now take ``$\approx$'' to simply mean the sum of the squares of the residuals over all pixels. A value of $l=0.5$ would approximate weighting by the uncertainty contributed by shot noise.  At $l=1$, each pixel is equally weighted, but this does not guarantee the smallest residual for each pixel. One advantage of our ePSF model is its high precision for faint stars, so we prioritize minimizing the residuals of dimmer pixels to enhance its performance on the faint end further. The median absolute deviation (MAD) of the residual is a natural choice of the goodness of fit to represent the majority (dimmer) pixels. Figure \ref{fig:weight} shows the normalized MAD of the residual versus different powers for 196 different cutouts sampled from Sector 7. We tested weights from $l = 0.4$ (prioritizing brighter pixels) to $l = 2$ (prioritizing dimmer pixels) and the smallest MAD happens near $l = 1.4$. We, therefore, adopt $l = 1.4$ in all fits. In addition, there are saturated pixels and dim vertical lines {seemingly corresponding to bad CCD columns} (different from the straps we discussed in Section \ref{sec:background}) in some FFIs. We assign these pixels $l = \infty$ (i.e.~zero weight) to exclude them from the fit. {Figure \ref{fig:epsf_examples} shows the resulting ePSF shapes from a part of Sector 1. The width of the ePSF gradually narrows down towards the upper-right of this CCD (the center of the camera lens) due to the TESS optical design. The trend shown is indeed gradual, so it is appropriate to assume the PSF is constant in each cutout. }

Despite the {535} parameters that we must fit, the linearity of the problem results in a modest computational cost. Deriving the ePSF for a single FFI cutout requires setting up and solving a single matrix equation, and takes {$\approx$ 0.4 seconds on a single core of a modern server processor \footnote{AMD EPYC 7313 16-Core Processor}}. Processing thousands of these cutouts {for each epoch}, a precondition to producing light curves, takes {$400$ seconds for the primary mission on a single core}.  {Extending this approach to all regions on a CCD ($\times 14^2$), and to all CCDs on the TESS camera ($\times 16$), brings the cost to about 350 CPU hours per primary mission sector. This remains computationally tractable on a modern server with tens of cores, requiring a wall-clock time of less than a day per primary mission sector ($\sim$ 3 days for the first extended mission and $\sim$ 9 days for the second extended mission).} In this way, the data may be reprocessed if needed. Deriving a local, free, non-parametric ePSF for every region of the TESS camera and every full-frame image is a linear problem, and thus a computationally tractable task.

The ePSF approach to photometry has a further advantage over fits using parametric formulae, e.g., a Gaussian or a Moffat model for the PSF.  A Gaussian or Moffat model requires the modeled positions of all stars on the TESS image to be correct in absolute terms so that the star's center corresponds to its peak intensity.  An ePSF approach has the weaker requirement that only the relative positions of the stars are correct.  If all stars are offset by a fraction of a pixel, then the modeled ePSF will have its peak flux correspondingly off-center.  There will be no consequence to the quality of the forward model or the extracted photometry of any star. 

\section{Extracting Light Curves} \label{sec:lc_extraction}
\subsection{PSF photometry light curves} \label{subsec:psf}
\begin{figure}
    \centering
    \includegraphics[width=\columnwidth]{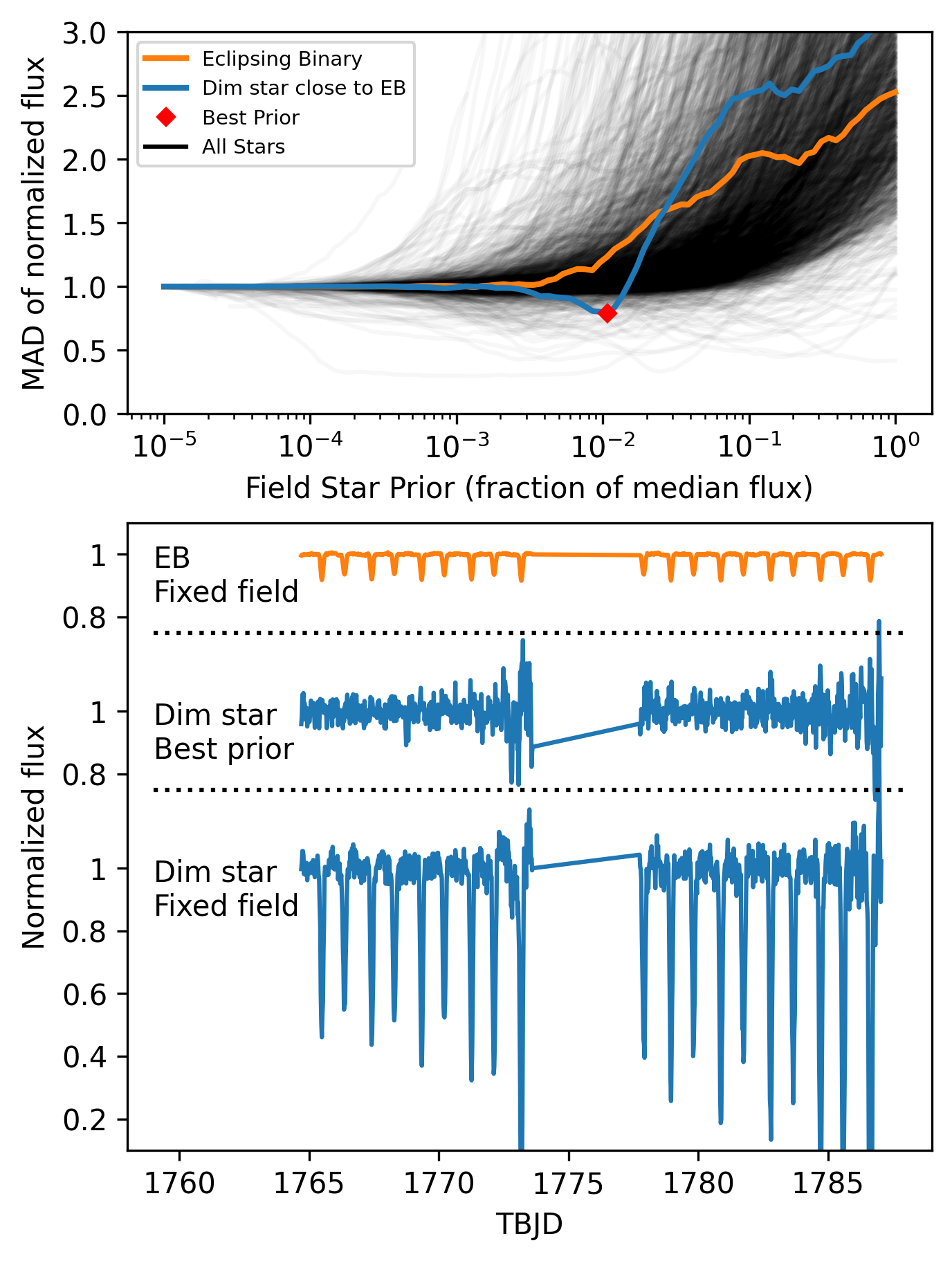}
    \caption{{Comparing the MADs of the PSF light curves if the field stars are assigned different priors. All curves are normalized to their MAD at prior = $10^{-5}$, effectively fixing the field stars to the Gaia-predicted flux. Top: The MAD curves of all Sector 17 stars (black) within 20 arcminutes of TIC 270022476 (an 11.5 TESS magnitude eclipsing binary (EB); orange). Gaia DR3 2015669353645091072 is highlighted in blue, a 15.2 TESS magnitude star 1 pixel away from the EB. The trend of all stars shows a strong preference towards fixing the field stars with only several exceptions. Bottom: The comparison among EB lightcurve with fixed field stars, dim star light curve fitted with fixed field stars, and dim star light curve fitted with the prior that returns the lowest MAD (prior = 0.01072; red diamond in the top panel). The dim star light curve with fixed field stars shows deep transits from the EB while allowing the field stars to float within the best prior removes most contamination. }}
  \label{fig:prior}
\end{figure}
Our final step after constructing the ePSF and establishing the background is to extract a light curve for each star. In previous sections, we described our ePSF fit by fixing the brightness of each star to its value measured by Gaia and converting it to the TESS photometric system. {With the fitted ePSF, we first explore the possibility of allowing all stars in the field to float while extracting the light curve. As discussed in Section \ref{sec:method}, we often have more stars than pixels in crowded fields, resulting in underconstrained fits. We try to resolve this problem by assigning Gaussian priors to field stars while fitting the target star. The priors are the fraction of each field star's median flux derived from Gaia, so using a very small prior fundamentally sets the field stars as non-variables. Figure \ref{fig:prior} shows the MADs of the PSF light curves when stars are fitted with different priors. The MADs of all stars from the vicinity of TIC 270022476 (an 11.5 TESS magnitude eclipsing binary (EB)) in Sector 17 (black) show an increasing trend as the prior gets wider, which supports choosing fixed field stars in general. Several exceptions could reach lower MADs at specific priors, and one of them is Gaia DR3 2015669353645091072 (blue). This dim star at 15.2 TESS magnitude is only 1 pixel away from the EB and is vulnerable to its transit contaminations if the field stars (including the EB) are considered fixed (the last row of the lower panel). If we allow the field stars to float with the best prior (the prior that results in the lowest MAD), most of the contamination is removed. Therefore, this approach is useful for decontaminating stars near variable sources if one finds the best prior by sampling. However, running this fit requires $\sim$8000 CPU hours for a single sector due to a large number of variables, not to mention that we have to sample the priors to find the one resulting in the largest decrease in MAD. It is thus impractical to fit all light curves with this method, especially when setting all stars fixed gives very high (if not the highest) precision at a very low computational cost. The function for sampling and applying priors is kept in the $\tglc$ package for manually decontaminating certain stars. }

We now allow the photometry of a given target star to float while holding the photometry of its neighbors fixed and keeping the ePSF derived earlier for the full $150 \times 150$ pixel region. This change is equivalent to modeling a FFI cutout for each cadence and then taking the residual of this from the measured cutout. {Since the field stars are fixed, the FFI cutout varies in time mainly due to background fluctuation.} We can then perform PSF photometry using the known ePSF on the residual images. This gives a perturbation to the flux derived from Gaia using the conversion of 15000 $\mathrm{e^-} / {\rm s}$ for a 10th {TESS} magnitude star given in the TESS Instrument Handbook v0.1. Adding this perturbation to the {Gaia-derived} flux gives our PSF photometry {flux} of the target star. In this way, extracting a light curve for each star in the field incurs a negligible computational cost over that for constructing the ePSF. 

\begin{figure*}
    \centering
    \includegraphics[width=\textwidth]{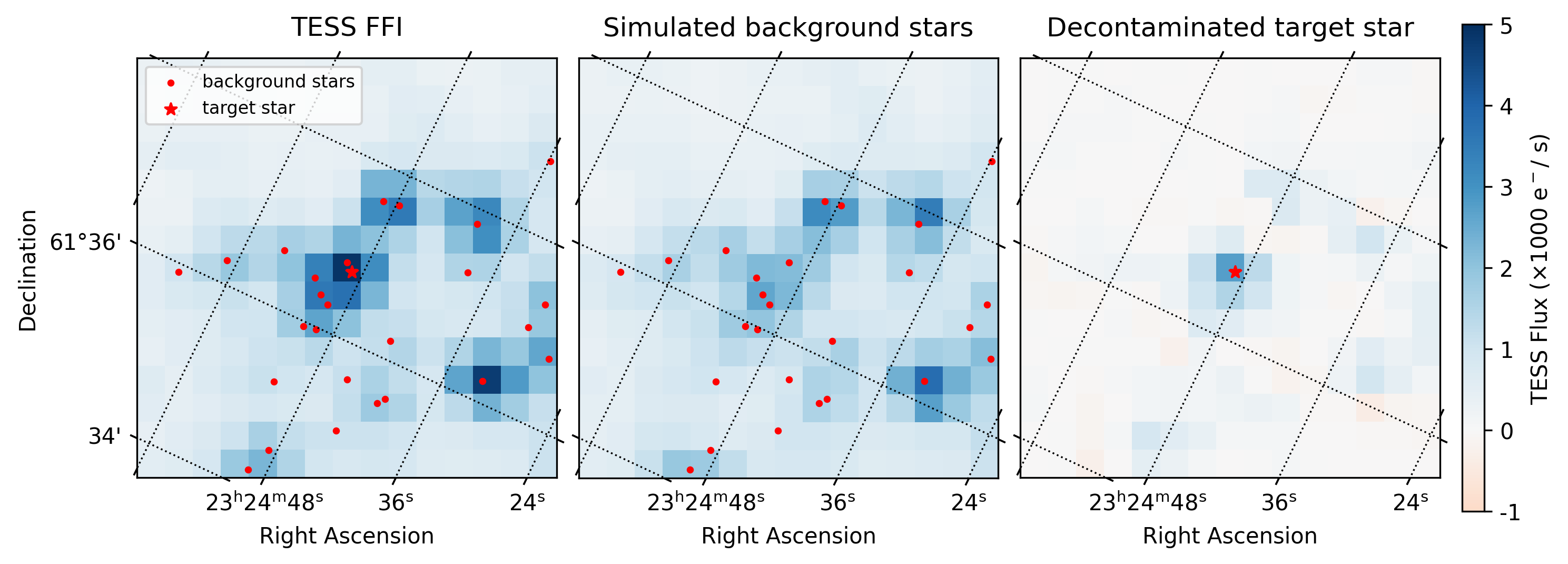}
    \caption{Illustration of an ePSF-based fit to a cutout of a TESS full-frame image. The left image is a cutout of the FFI; the middle is our forward modeling of all stars except the target star (10.5 TESS magnitude); the right image is the residual image as the difference between the first two images. The residual image is decontaminated and is ready to generate light curves. {The unevenness in the residual image is a combined result of the imperfections of our modeling, such as constant field stars, ePSF shape, and spatially variable background. }}
  \label{fig:ePSF_example}
\end{figure*}

Figure \ref{fig:ePSF_example} illustrates the process of extracting a light curve. The left-hand panel shows a small cutout of a TESS full-frame image, a subset of that used to build the ePSF model. The middle panel shows the results of a forward model of the image fixing all stars to their Gaia-inferred photometry; we have omitted the central target star from this model. The right-hand panel shows the residual from the model subimage after removing the neighboring stars (but not removing any modeled light from the target star). Performing PSF photometry on this image will produce one data point in the light curve of the target. The same can be done for all frames and all stars in this cutout to generate light curves from PSF photometry.

\subsection{Aperture photometry light curves} \label{sec:aper}
\begin{figure*}
    \centering
    \includegraphics[width=\textwidth]{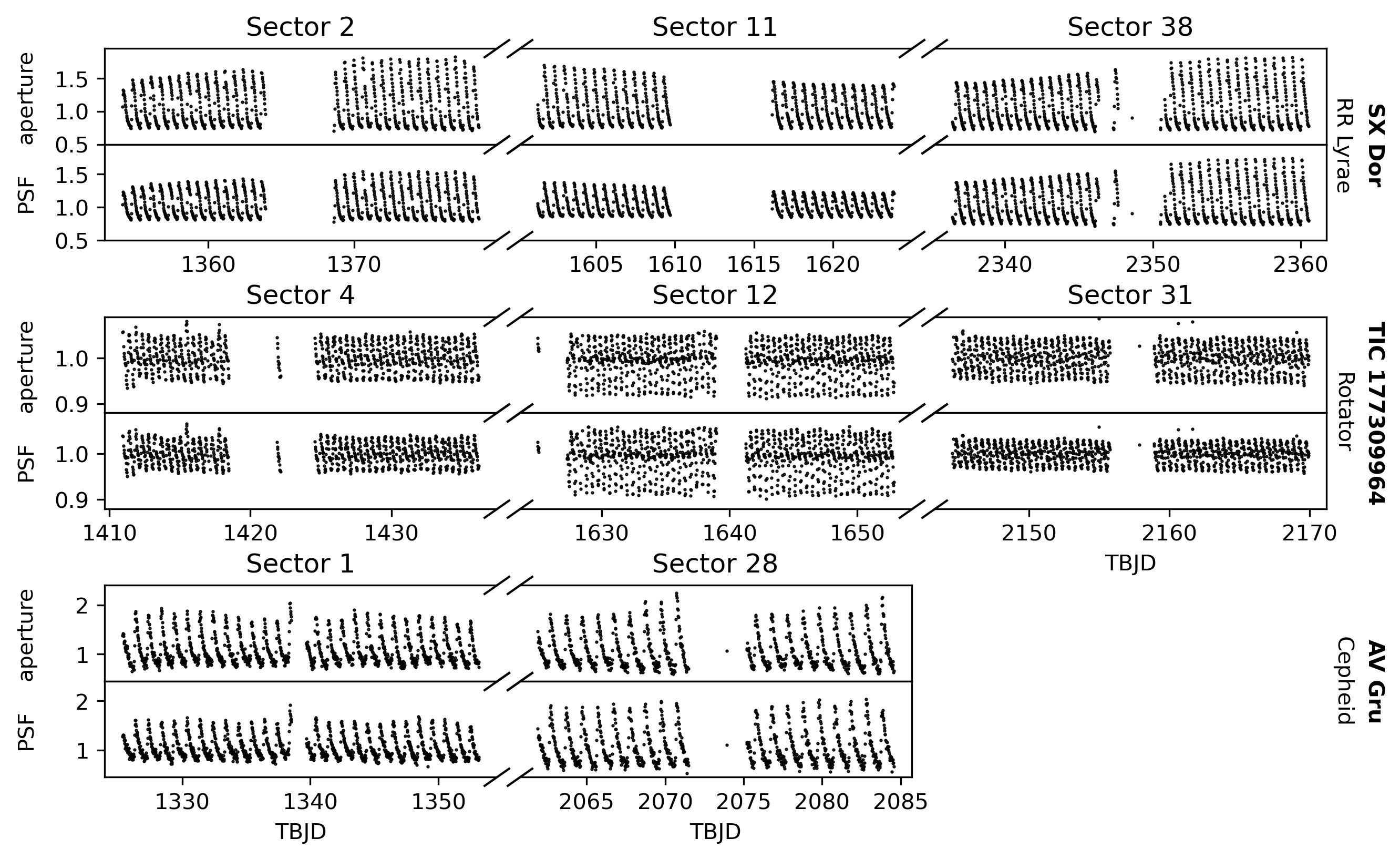}
    \caption{TGLC light curves of 3 different type variable stars. TOP: SX Dor is an RR Lyrae star near the outskirts of the LMC and thus a difficult target to deblend. MID: TIC 177309964 is a rapid rotator. BOT: AV Gru is a faint Cepheid variable with a TESS magnitude of 16.94. TGLC Aperture light curves offer more reliable amplitude estimations for these highly variable stars in a crowded field than TGLC PSF light curves. {Note: These light curves are detrended with \textsf{wotan} with a window length of 1 day (the default for all calibrated TGLC light curves). Users should consider the non-calibrated version if dealing with long-period variable stars because such signals could be removed by detrending. }} 
   \label{fig:variables}
\end{figure*}

We also generate TGLC Aperture light curves. Using the residual image described in Section \ref{subsec:psf}, it is straightforward to obtain a photometric data point by summing pixels within an arbitrary aperture shape. The optimal choice of aperture is a non-trivial problem for each star. We allow manual extraction of the time series from the reduced images in our package $\tglc$. It is possible to choose the aperture and produce customized aperture light curves. We provide an empirical arbitrary choice of $3\times3$ aperture to produce the default aperture light curves published together with the PSF light curves. Since only a part of the light from a star falls onto this aperture, the background levels of the aperture light curves need to be corrected. We estimate the median total flux of each star based on Gaia and calculate the percentage of light that shall fall onto the $3\times3$ pixels region based on the ePSF shape. We then shift the aperture light curve's median to the Gaia-predicted median multiplied by this percentage. 

While the PSF light curves have consistently high quality, aperture photometry has an edge in most conditions. Figure \ref{fig:variables} shows three variable stars of different types: SX Dor is a RR Lyrae star near the outskirts of the Large Magellanic Cloud and is considered challenging to deblend by \cite{rr_lyrae}; TIC 177309964 is a rapid rotator with a period of 0.4533 days \citep{rotator}; AV Gru is a faint Cepheid star with a TESS magnitude of 16.94 \citep{cepheid}. The amplitudes of all three stars vary considerably in each cycle, but aperture light curves are more consistent across sectors than PSF light curves. The PSF light curves' inconsistent amplitudes result from their vulnerability to imperfect decontamination. {As the target star gets dimmer, the remaining contamination in the residual image becomes more noticeable.} These constants are added to the pixels' function as `anchor points' to drag the PSF fit closer to their value. This generally reduces the amplitude of the PSF light curve variations, like the Sector 11 PSF light curve of SX Dor. Removing this artifact requires better decontamination and is thus a non-trivial task. However, the remaining constants do not affect the aperture light curve because they only affect the background level which can be easily removed as discussed in the previous paragraph. Therefore, aperture light curves have a more reliable amplitude estimation for crowded fields and dim stars. 

\subsection{Weighted light curves}

{The previous two sections described two different methods of deriving TESS FFI light curves.  Each will have slightly different measurement errors. The pixels used are the same, so the instrumental noise is shared, but the weighting of the individual pixels differs in the two approaches.  As a result, we expect the measurement errors to be significantly, but not perfectly, correlated.  A linear combination of the two light curves---aperture and PSF---should have lower noise than either light curve.  To explore this possibility, we introduce TGLC Weighted light curves.}

{TGLC Weighted light curve is the weighted average of TGLC PSF and TGLC Aperture light curve. In several test fields, we find that the weighted average of the two versions of the light curve can subtly increase photometric precision over both versions. The optimal weights to produce the highest precision light curves are field-dependent, mostly correlated with the crowdedness of the field. We lack an analytical solution for each star's best weight, so sampling is necessary for the best result. The TGLC Weighted light curves are not published with the PSF and aperture light curves, but they can be derived from these two versions with an arbitrary choice of weights. An example of the TGLC Weighted light curve precision improvement is shown in Section \ref{sec:comparisons}.
}

\section{Light Curve Comparison} \label{sec:comparisons}
\begin{figure*}
    \centering  
    \includegraphics[width=\textwidth]{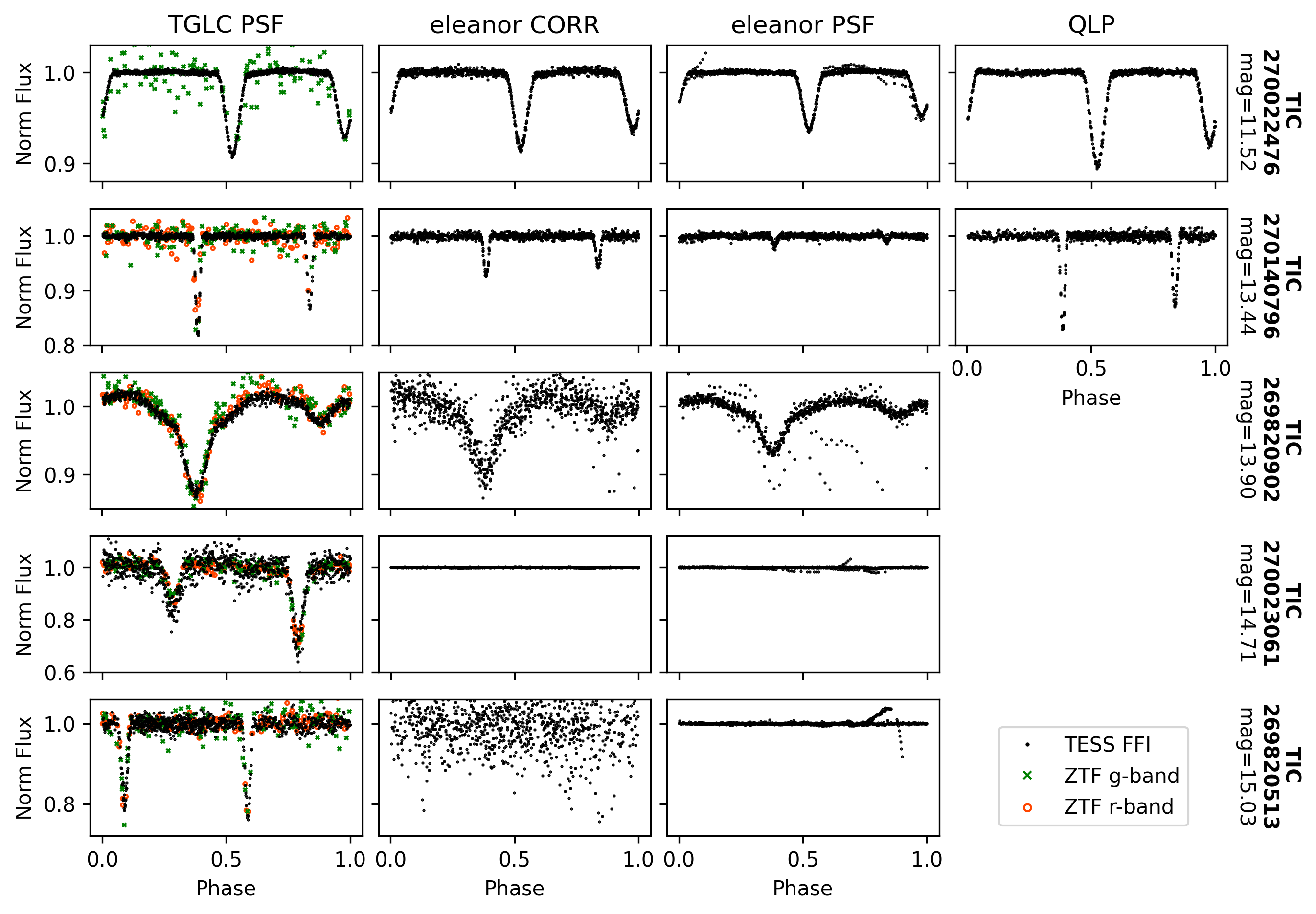}
    \caption{Comparison of TGLC PSF light curves {(first column)} with {\tt eleanor} \citep{eleanor} {(second and third column)} and QLP \citep{QLP} databases {(last column)} for five sample eclipsing binaries. {All light curves are detrended with \textsf{wotan} \citep{wotan}.} Stars range in brightness from 11.5 {TESS} magnitude (top) to 15th {TESS} magnitude (bottom). We omit TGLC Aperture light curves because they are indistinguishable from TGLC PSF light curves by eye. Phase-folded Zwicky Transient Facility light curves \citep{ZTF} confirm the eclipse depths shown in the left column. Quality drops rapidly for other pipelines when the target gets dimmer while TGLC PSF has consistently high precision. Note that for both {\tt eleanor} light curves for TIC 270023061, the transits are detected, but at a very shallow depth imperceptible in the figure. }   \label{fig:eb}
\end{figure*}

\begin{figure*}
    \centering  
    \includegraphics[width=\linewidth]{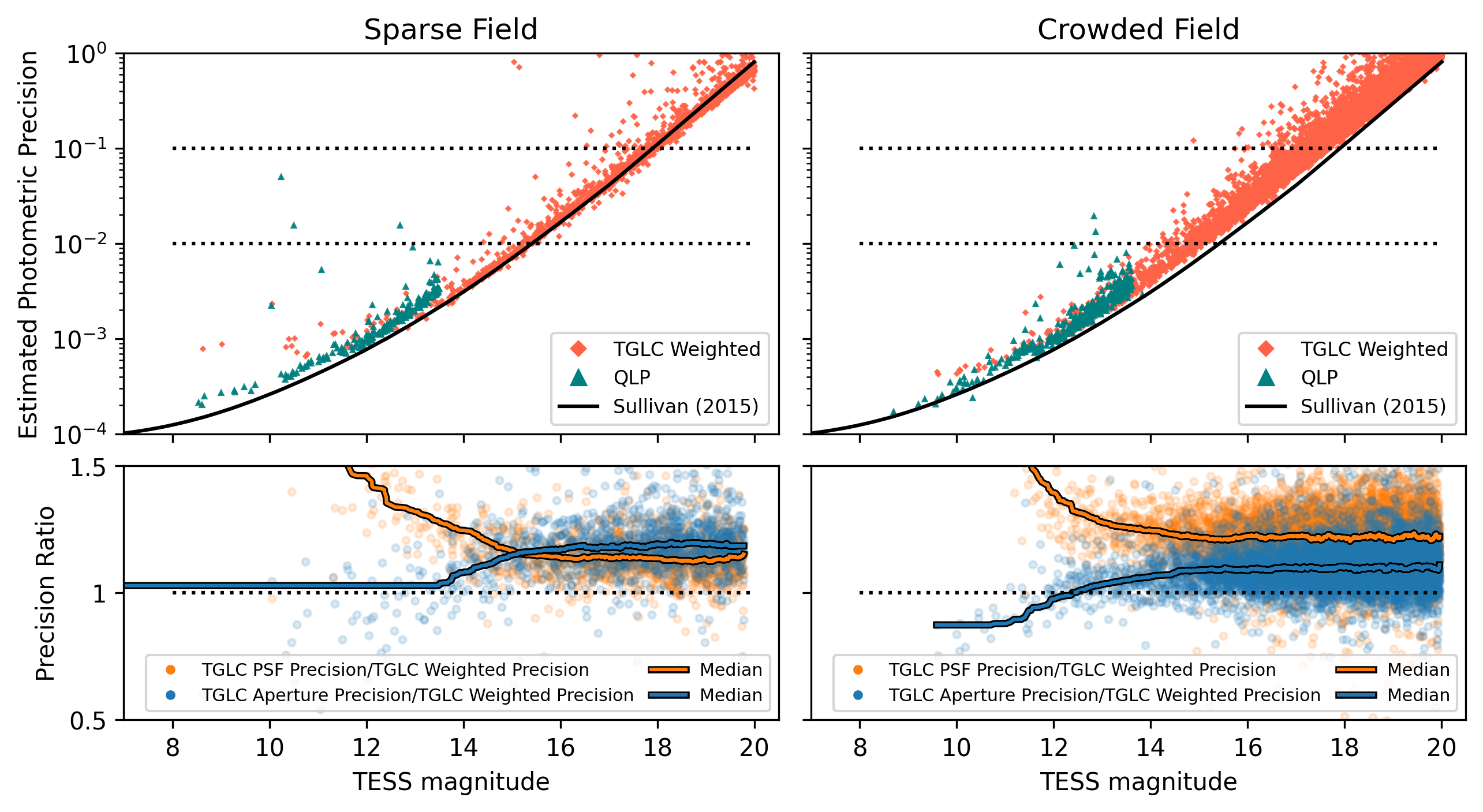}
    \caption{{The photometric precision (Equation \ref{eq:precision}) of TGLC compared to the pre-launch prediction \citep{old_ffi_predictions} in both sparse (left column) and crowded field (right column). The highest point-to-point precision is achieved by a certain weighted average of TGLC PSF and TGLC Aperture; The shown TGLC Weighted is the photometric precision of 0.4PSF + 0.6Aperture. Top: The TGLC Weighted light curve precision (orange diamond) closely tracks the pre-launch prediction (black line) even in the crowded field. The QLP (green triangles) precision is very close to the pre-launch prediction in the brighter end. Bottom: The precision ratio of TGLC PSF and Aperture versus TGLC Weighted. The TGLC PSF is possible to outperform TGLC Aperture in sparse field for dim stars (the smaller the ratio is, the better), but TGLC Aperture has higher precision in other cases. }}
   \label{fig:precision}
\end{figure*}

In this section, we compare the light curves from TGLC to those of the QLP \citep{QLP} and {\tt eleanor} \citep{eleanor}.  We begin with the light curves of five sample EBs that are all found in Sector 17 near NGC 7654.  Figure \ref{fig:eb} shows our TGLC PSF together with those of {\tt eleanor} \citep{eleanor} and QLP \citep{QLP}, as well as Zwicky Transient Facility (ZTF) light curve \citep{ZTF}. ZTF has much better spatial resolution than TESS, greatly reducing the need to deblend in this moderately crowded field.  

For all five EBs, TGLC PSF shows the lowest noise and the best agreement with ZTF. At a {TESS} magnitude of 11.5, an expected comfort zone for all pipelines, QLP performs well but {\tt eleanor} struggles to deblend the photometry without modeling neighboring stars' PSFs. The resulting inconsistency in {\tt eleanor} is more prominent for dimmer stars: this is caused by a systematic lack in removing the contamination. An overestimate of the contamination could deepen the transit and vice versa. QLP, with its difference imaging approach anchored to the TESS Input Catalog photometry, is relatively better for this issue, but it cuts off at 13.5 {TESS} magnitude. TGLC PSF also shows a smaller point-to-point scatter in its photometry, i.e., a higher photometric precision. Finally, our ePSF background fit value enables us to implement quality flags that remove outliers accurately and automatically (light curves of other pipelines may also be filtered by their quality flags). We label the frames that have background values 5 standard deviations away from the median as bad frames in the quality flag. These sudden background increases are usually caused by scattered light from the Earth and the Moon at the ends of each $\sim$14 day observation window. It is very hard to recover those irregularly contaminated frames, so we label them as low-quality frames {with TGLC flag (see Appendix \ref{appendix:a3})}. 

{We next systematically analyze the photometric precision of TGLC light curves by comparing all stars in two $35' \times 35'$ regions in Sector 7 and 17. The Sector 17 field is relatively sparse and has an average of 0.2 Gaia stars (all magnitudes) per pixel; the Sector 7 field is crowded with 1.2 Gaia stars per pixel on average.  We assess our photometric precision using a robust estimate of the point-to-point scatter in photometry. We first take the flux differences between adjacent fluxes ($D$). The root-mean-square of this difference is vulnerable to variable sources and outliers, so we adopt the median absolute deviation and multiply it by a factor of 1.48 (the ratio of the median absolute deviation and the standard deviation for Gaussian errors). We then divide the point-to-point scatter by $\sqrt{2}$ to estimate the photometric precision of each point rather than on the difference between two points. The estimated photometric precision for each light curve is then
\begin{equation}\label{eq:precision}
    \text{precision} = \frac{1.48}{\sqrt{2}} \text{ median}(|D|).
\end{equation}
All of our light curves and estimated precisions refer to a single 30-minute photometric point.}

{The top row of Figure \ref{fig:precision} shows the comparison among TGLC Weighted (orange diamonds), QLP (green triangles), and an aperture photometry prediction from \cite{old_ffi_predictions} (black line). We empirically choose TGLC Weighted = 0.4 TGLC PSF + 0.6 TGLC Aperture for this analysis since these weights result in the highest precision in these fields. We derive the pre-launch prediction from Figure 14 of \cite{old_ffi_predictions}, which summarizes the base noise level of TESS photometry. In both fields, the TGLC Weighted precision almost reaches the base noise level and displays a narrow distribution, indicating good control over systematics. QLP light curve precision closely tracks the pre-launch prediction until 13.5 TESS magnitude. 
We do not show {\tt eleanor} on these plots.  Its precision in a crowded field surpasses that in a sparse field. This indicates significant uncorrected dilution, where flux from neighboring stars and backgrounds contributes to a light curve (c.f.~ the flat {\tt eleanor} light curves of TIC 270023061 in Figure \ref{fig:eb}).  This dilution can be mitigated with user involvement, but it prevents straightforward comparisons of precision across fields. The bottom row of Figure \ref{fig:precision} compares the precision of TGLC PSF to the precision of TGLC Aperture normalized by the precision of TGLC Weighted. The TGLC Aperture has higher precision than the TGLC PSF in general (examples shown in Section \ref{sec:exoplanet}), but TGLC PSF has an edge in sparse fields for stars dimmer than 15 TESS magnitude. The precision improvement of TGLC Weighted over the other two TGLC light curves is $\sim$20\%. All three TGLC light curves reliably achieve a precision of $\lesssim$2\% for 16th {TESS} magnitude stars in 30-minute data. 
}

\section{Exoplanet light curves case study} \label{sec:exoplanet}

\begin{figure*}
    \centering
    \includegraphics[width=\textwidth]{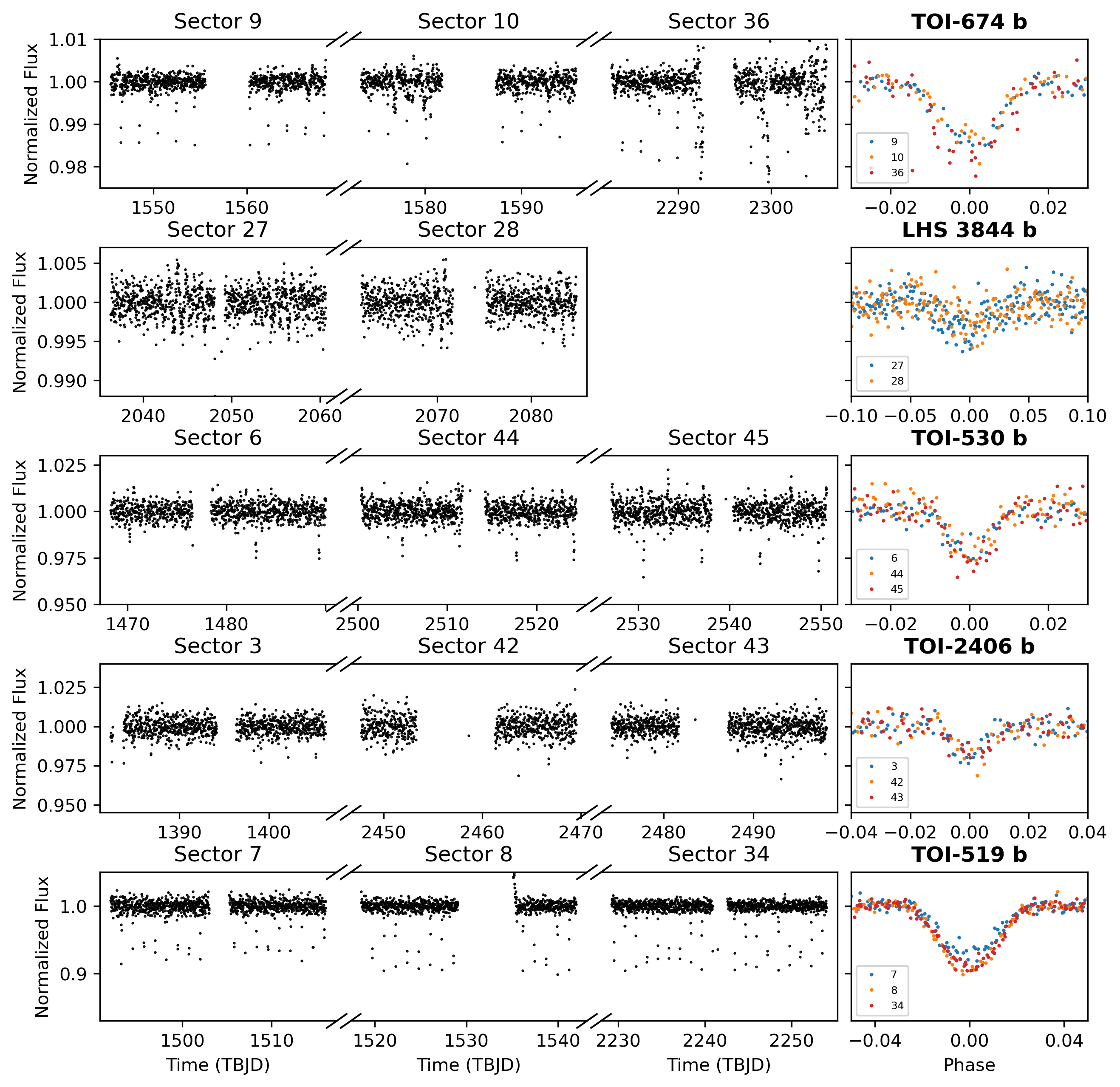}
    \caption{TGLC PSF light curves of 5 exoplanets in multiple sectors. Light curves of the same exoplanets from existing pipelines are shown in Figure \ref{fig:exoplanet_other}. These light curves show higher precision than FFI light curves from other pipelines. Phase fold periods are adopted from Table \ref{tab:exoplanet} TGLC PSF periods. Note that for TOI-519 b, the PSF light curve has unreliable transit depths across sectors due to the same reason discussed in Section \ref{sec:aper}} 
   \label{fig:exoplanet_psf}
\end{figure*}
\begin{figure*}
    \centering
    \includegraphics[width=\textwidth]{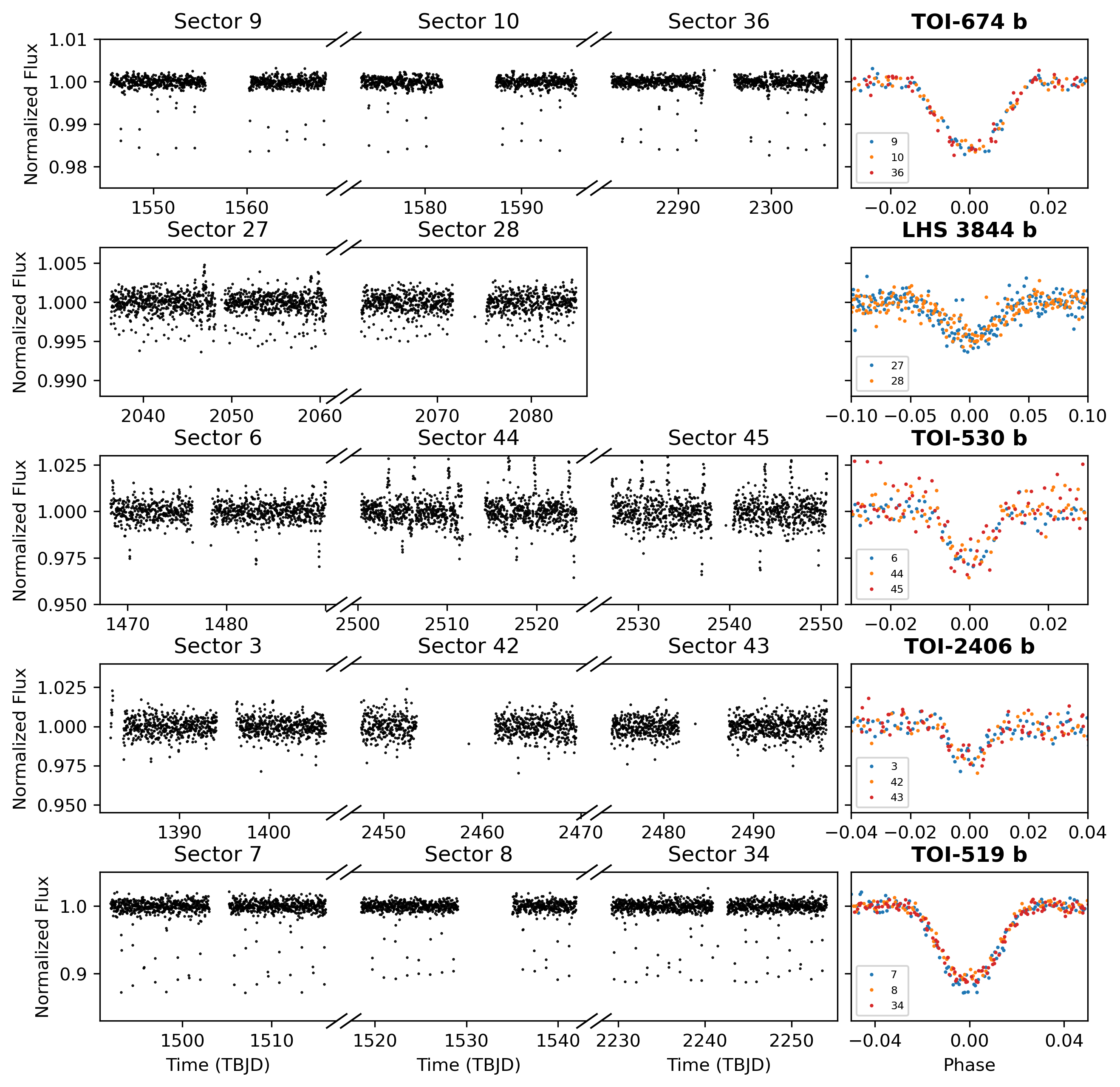}
    \caption{TGLC Aperture light curves of 5 exoplanets (same as those in Figure \ref{fig:exoplanet_psf}) in multiple sectors.}
   \label{fig:exoplanet_aper}
\end{figure*}

\begin{figure*}
    \centering
    \includegraphics[width=\textwidth]{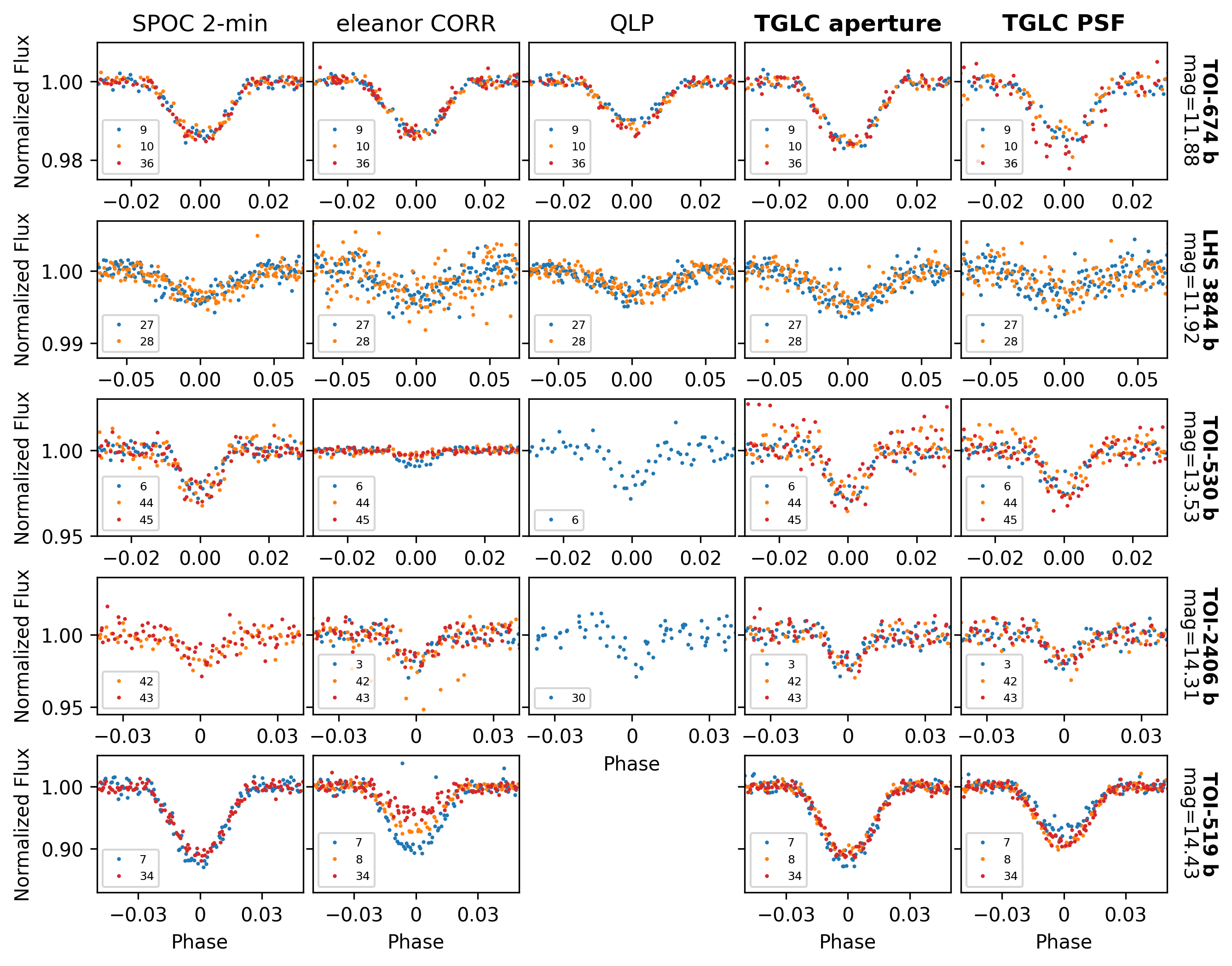}
    \caption{TGLC PSF light curves of five exoplanets in multiple sectors. Light curves of the same exoplanets of existing pipelines are shown in Figure \ref{fig:exoplanet_other}. TGLCs have much higher precisions than FFI light curves from other pipelines. Phase fold periods are adopted from Table \ref{tab:exoplanet}  TGLC PSF periods. } 
   \label{fig:exoplanet_all}
\end{figure*}

We perform another case study of TGLC, analyzing its light curves for five exoplanets from TESS's primary mission. $90\%$ of confirmed TESS exoplanet discoveries are from stars brighter than 12th {TESS} magnitude\footnote{https://tess.mit.edu/publications/}, but there are vastly more potential exoplanet hosts dimmer than 12th {TESS} magnitude.Therefore, we choose five of the faintest TESS exoplanet hosts for our case study, ranging in {TESS} magnitude from 11.9 to 14.3. The analysis of these examples demonstrates the potential of TGLC to enable the discovery and characterization of new exoplanets. 

Figure \ref{fig:exoplanet_psf} and Figure \ref{fig:exoplanet_aper} show five exoplanets' TGLC PSF and TGLC Aperture light curves respectively. All light curves are binned to a 30-min cadence (for the extended mission) and are detrended with \textsf{wotan} \citep{wotan}. Other pipelines' light curves for the same exoplanets are shown in Figure \ref{fig:exoplanet_other}. Both TGLC PSF and TGLC Aperture show excellent results compared to other FFI pipelines (Figure \ref{fig:exoplanet_all}). Up to 3 different sectors of observations of each star are distinguished in the phase-folded plots to show cross-sector consistency. Both methods show steady low noise levels and consistent transit shapes for all five exoplanets. It is challenging to maintain consistent light curves between sectors or even between two observation windows in a single sector: for example, {\tt eleanor} PSF in Figure \ref{fig:exoplanet_other} has inconsistent transit depth between sectors for TOI-674 b and within a sector for TOI-519 b. {Spacecraft direction changes between sectors and rotates the field of view, which shifts and rotates all stars. The contaminations are thus different in each sector for the same star and require independent modeling.} Only if the decontamination method is robust against different contaminations can consistent results be produced in different sectors. TGLC fully models each frame independently from Gaia priors, so it can model the change of contamination between sectors properly.  

We then use {\tt exoplanet} \citep{exoplanet:joss} to model our light curves and the results are shown in Table \ref{tab:exoplanet}. These light curve fits are kept to their simplest form with only the necessary free parameters {(rows of Table \ref{tab:exoplanet})}. We compare our fitted values of the most important orbit parameters, stellar and planet radii, and stellar masses to the published values. Four exoplanets' period fits are improved because of the extended time baseline with new sectors except LHS 3844 b, which has an ultra-short period of $\approx 0.46$ days. The FFI light curves with a cadence of 10-min in Sector 27 and Sector 28 are a bit sparse for the period to converge; in contrast, SPOC 2-min light curves can be fitted well with the same priors. The relatively longer cadence is an intrinsic disadvantage for FFI light curves, so we set the periods and reference transit time equal to published values from \cite{Vanderspek_2019} for LHS 3844 b. Our fits mostly agree with the published values within 2 standard deviations for the other fitted parameters. Since our fit is a lone light curve fit, we do not expect it to outperform the publication fits that use multiple instruments' data in general. However, combined with radial velocity measurement and ground-based follow-up photometric measurement, a comprehensive TGLC exoplanet fit could improve the precision of all free parameters. 

Our method has its limits. Comparing two TGLC light curves for TOI-519 b shows a percent-level discrepancy in transit depth for TGLC PSF. As we discussed in Section \ref{sec:comparisons}, TGLC PSF can achieve $\lesssim$2\% photometric precision for a 16th {TESS} magnitude star; TOI-519 is 14.4 {TESS} magnitude so we can expect a slightly better precision. The reason for this inconsistency is the same as the variable star case we discussed in Section \ref{sec:aper}. It is also worth mentioning that the TGLC PSF photometry for Sector 44 and Sector 45 of TOI-530 b (Figure \ref{fig:exoplanet_psf}) has much fewer spikes than we see in the TGLC Aperture (Figure \ref{fig:exoplanet_aper}). TGLC PSF deblends better in the presence of nearby variable sources. Since PSF and aperture light curves both have their advantage in certain scenarios, we publish both in our data release. 

\begin{deluxetable*}{lcccccc}
\tablecaption{Exoplanet modeling}
\tablewidth{0pt}
\tablehead{Identifier & TOI  674 b & LHS 3844 b\tablenotemark{a}  & TOI 530 b & TOI 2406 b & TOI 519 b & Ref.}
\startdata
\begin{tabular}[c]{@{}l@{}}TESS \\ magnitude\end{tabular} & 11.8764 & 11.9238 & 13.5287 & 14.3109 & 14.4347 & Gaia DR3\tablenotemark{b} \\ \hline
\multirow{3}{*}{\begin{tabular}[c]{@{}l@{}}Period \\ (days)\end{tabular}} & $1.97716[5]$ & \multirow{3}{*}{$0.46292[913]$} & $6.3875[83]$ & $3.0766[76] $ & $1.265232[0]$ & TGLC PSF \\
 & $1.97716[2]$ &  & $6.3875[82] $ & $3.0766[15] $ & $1.265233[0]$ & TGLC Aper. \\
 & $1.97714[3]$ &  & $6.3875[97]$ & $3.07668[96] $ & $1.265232[8]$ & Literature \\ \hline
\multirow{3}{*}{\begin{tabular}[c]{@{}l@{}}$\mathrm{T_c}$ \\ (TBJD)\end{tabular}} & $1546.501[4]$ & \multirow{3}{*}{$1325.725[58]$} & $1470.20[20]$ & $1383.72[33]$ & $1493.142[45]$ & TGLC PSF \\
 & $1546.502[0]$ &  & $1470.20[24]$ & $1383.3[23]$ & $1493.142[39]$ & TGLC Aper. \\
 & $1546.501[7]$ &  & $1470.19[98]$ & $1383.723[35]$ & $1493.142[35]$ & Literature \\ \hline
\multirow{3}{*}{b} & $0.5 \pm 0.2$ & $0.15 \pm 0.10$ & $0.28 \pm 0.18$ & $0.14 \pm 0.08$ & $0.25 \pm 0.15$ & TGLC PSF \\
 & $0.4 \pm 0.2$ & $0.15 \pm 0.10$ & $0.25 \pm 0.18$ & $0.17 \pm 0.09$ & $0.22 \pm 0.13$ & TGLC Aper. \\
 & $0.624 \pm 0.035$ & $0.186 \pm 0.064$ & $0.33 \substack{+0.08 \\ -0.11}$ & $0.16 \substack{+0.15 \\ -0.11}$ & $0.19 \substack{+0.06 \\ -0.09}$ & Literature \\ \hline
\multirow{3}{*}{\begin{tabular}[c]{@{}l@{}}$\mathrm{R_{planet}}$ \\ $(\mathrm{R}_{\oplus})$\end{tabular}} & $5.2 \pm 0.3$ & $1.32 \pm 0.04$ & $8.2 \pm 0.6$ & $2.85 \pm 0.16$ & $11.1 \pm 0.6$ & TGLC PSF \\
 & $5.5 \pm 0.2$ & $1.41 \pm 0.04$ & $9.0 \pm 0.6$ & $2.30 \pm 0.15$ & $12.7 \pm 0.5$ & TGLC Aper. \\
 & $5.25 \pm 0.17$ & $1.303 \pm 0.022$ & $9.3 \pm 0.7$ & $2.94 \substack{+0.17 \\ -0.16}$ & $8.4 \pm 2.4$ & Literature \\ \hline
\begin{tabular}[c]{@{}l@{}}Literature\\ Ref.\end{tabular} &  \begin{tabular}[c]{p{0.25\columnwidth}l} \cite{Murgas_2021} \end{tabular} & \begin{tabular}[c]{p{0.25\columnwidth}l}\cite{Vanderspek_2019} \end{tabular} & \begin{tabular}[c]{p{0.25\columnwidth}l}\cite{Gan_2021} \end{tabular} & \begin{tabular}[c]{p{0.25\columnwidth}l}\cite{Wells_2021} \end{tabular} & \begin{tabular}[c]{p{0.25\columnwidth}l}\cite{Parviainen_2020}\end{tabular}&     
\enddata
\tablenotetext{a}{Due to its extremely short period, the 10-min cadence FFI data cannot fit LHS 3844's period well. We fixed the period and reference transit time using the literature value for this fit. }
\tablenotetext{b}{Gaia DR3 magnitude converted to the TESS band using the relations from the TESS Instrument Handbook  v0.1}
\label{tab:exoplanet}
\end{deluxetable*}

\section{Data and availability} \label{sec:data}

All our TGLC data products are available at MAST as a High Level Science Product via \dataset[10.17909/610m-9474]{\doi{10.17909/610m-9474}}. {The primary mission light curves are released with the paper via bulk download, and the first extended mission sectors of TGLC are continuously produced. The ingesting process for the MAST portal query takes longer than bulk download, but new sectors are updated weekly.} As the second extended mission sectors are available, we will continue delivering new light curves. We cut each FFI ($2048 \times 2048$ pixels) into $14 \times 14$ cutouts, each with $150 \times 150$ pixels. This leaves two-pixel-wide overlaps between cutouts to keep most stars at least 2 pixels away from the edge. Each cutout is then passed to the ePSF model and background model to calculate the best fit ePSF. We then produce light curves of all stars brighter than 16th TESS magnitude, and each file includes four light curves: a PSF light curve, an aperture light curve, and their calibrated versions. The calibrated light curves are detrended and normalized with \textsf{wotan} \citep{wotan}. {The format of the light curve FITS file is detailed in Appendix \ref{appendix}.} The package $\tglc$\footnote{\url{https://doi.org/10.5281/zenodo.7023845}} is pip-installable\footnote{\url{https://pypi.org/project/tglc/}} and offers more customized options for light curve fitting. It is best used for a small cut ($< 100 \times 100$ pixels) of the sky and multi-sector comparison. The user can get light curves for any star from released sectors with comparable precision to the MAST-released light curves. One can also choose to save a decontaminated image like the last panel in Figure \ref{fig:ePSF_example} for customized light curve extraction.

\section{Discussion} \label{sec:discussion}

TESS-Gaia Light Curve achieves {the photometric precision close to the instrumental noise level by incorporating the position and brightness measurements of Gaia DR3 in an effective PSF fit of TESS FFI. The photometric performance that we demonstrate in Section \ref{sec:comparisons} meets the noise levels assumed in predictions of TESS yields \cite{old_ffi_predictions}.} TESS full-frame images are expected to result in the discovery of thousands of transiting exoplanets, including $\sim$1000 around stars fainter than 12th {TESS} magnitude \citep{ffi_predictions}. These predicted discoveries can be realized with the improvements in photometric precision such as those provided by TGLC. 

TESS full-frame images are yielding significant scientific results despite the limitations in data precision and availability. Cataclysmic variable light curves and supernova light curves may be derived with high precision from full-frame images \citep{TESS_AMCVn,supernovae_1,supernovae}. Planet searches and eclipsing binary searches have been conducted on full-frame images, but on a limited scale \citep{CDIPS,PATHOS}. Other studies have had results limited by precision: \cite{fullframe_pulsators} discovered 28 subdwarf B stars in the southern TESS full-frame images, mostly around 14-16 {TESS} magnitude stars, but were able to identify asteroseismic pulsations in only two of them. These {14-16 {TESS} magnitude stars} are precisely the ones where we achieve the largest improvements over existing pipelines. A follow-up study searching for eclipsing binaries and pulsating stars was further limited by crowded fields and blending \citep{fullframe_pulsators2}. Our TESS-Gaia light curves overcome most limitations of blended stars down to 16th {TESS} magnitude. TGLC can open new horizons for TESS time-domain sciences and large-scale automated search for new periodic signals. 

TGLC still has several limitations that we will work to overcome in the future.  The first is the possibility of further variations in the background level at a star's location. TESS is subject to strong spatially variable backgrounds from scattered light. We will therefore measure whether a target star's flux relative to the median of its neighbors matches this ratio as observed by Gaia.  If the star is brighter or fainter than expected, it could point to an under-estimated or over-estimated background, respectively.  We will determine whether such a correction is needed and if so, to apply it to our light curves. 

The second limitation of TGLC is in deblending. Variable targets are still partially contaminating all stars nearby because our published light curves assume background stars to have constant flux. Fully deblending requires allowing all of a star's immediate neighbors to have variable fluxes. {We attempt this in Section \ref{subsec:psf} by assigning priors to field stars, which achieves better deblending at a large computational cost. With the future release of Gaia, we may only allow the most variable field stars to float and keep the number of free parameters under a reasonable number to improve performance. }  

With the release of Gaia DR3, including individual photometric time series in the vicinity of the Andromeda Galaxy, we plan to check our time series photometry against individual Gaia measurements.  Gaia DR4 lacks an expected release date, but it will include thousands of photometric data points for nearly every star brighter than 20 {TESS} magnitude.  These light curves will form a coarsely sampled, but precise, check on the TESS photometry.  They will serve as a verification of the deblending performed by the TESS-Gaia pipeline.

\begin{acknowledgments}
We are grateful to Ben Montet and Chelsea Huang for their suggestions regarding the development of our method. We acknowledge James Davenport for his early inspiration. We are thankful to Hannah M. Lewis and Scott W. Fleming for their help with data publication on MAST. We thank Corey Beard for helping with the {\tt exoplanet} fitting. We appreciate Aomawa Shields and Paul Robertson for their comments on this paper. We thank Mirek Brandt for his help in inspecting documentation for $\tglc$. We value the conversation about FFI WCS with Clara E. Brasseur. We are grateful for the revision advice from the anonymous reviewer.  T.D.B.~gratefully acknowledges support from the Heising-Simons Foundation under grant \#2019-1493 and from the Alfred P.~Sloan Foundation.
\end{acknowledgments}

Our pipeline uses numpy \citep{numpy}, scipy \citep{scipy}, astropy \citep{exoplanet:astropy13, exoplanet:astropy18}, and astroquery \citep{astroquery}. This research made use of {\tt exoplanet}\citep{exoplanet:joss,exoplanet:zenodo} and its dependencies \citep{celerite1,celerite2, exoplanet:agol20, exoplanet:arviz,exoplanet:astropy13, exoplanet:astropy18, exoplanet:kipping13,exoplanet:kipping13b, exoplanet:luger18, exoplanet:pymc3, exoplanet:theano}. 

\appendix
\section{TGLC data product description}\label{appendix}
The TESS Gaia Light Curves (TGLC) are published in MAST as a High Level Science Product (HLSP). The primary mission light curves are published with the paper and the following light curves are continuously produced. We follow the standard TESS light curve FITS file convention and make necessary adjustments. We describe the format of our data product in this appendix to help users utilize them efficiently. The most up-to-date information about the data product can be found in TGLC GitHub repository\footnote{\url{https://github.com/TeHanHunter/TESS_Gaia_Light_Curve}}. 
\subsection{File format}\label{appendix:a1}
TGLC FITS files follow the naming convention of HLSP:

hlsp\_tglc\_tess\_ffi\_gaiaid-\{Gaia DR3 ID\}-s\{sector number\}-cam\{camera number\}-ccd\{CCD number\}\_tess\_v1\_llc.fits

Each FITS file has two Header Data Units (HDUs). The primary HDU is only used when generating light curves with the option save\_aper=True when running $\tglc$. All light curves on HLSP have empty primary HDU. The secondary HDU includes the light curves in a binary table. 
\subsection{Light curve headers}\label{appendix:a2}
The primary header includes the Gaia measurements of the star and the TESS FFI information. The secondary header includes uncertainties of the light curves and other PSF fit parameters. Table \ref{tab:1header} and \ref{tab:2header} are the headers of TOI-519 b Sector 7 light curve. 
\startlongtable
\begin{deluxetable}{llcl}
\tablecaption{Primary Headers \label{tab:1header}}
\tablewidth{0pt}
\tablehead{\colhead{Header Card} &\colhead{Default/Example Value} &\colhead{Data Type} & \colhead{Description} }
\startdata
SIMPLE & True & bool & conforms to FITS standard \\
BITPIX & 8 & int & 8 / array data type \\
NAXIS & 0 & int & 0 / number of array dimensions \\
EXTEND & True & bool &  \\
NEXTEND & 1 & int & number of standard extensions \\
EXTNAME & `PRIMARY' & str & name of extension \\
EXTDATA & `aperture' & str & decontaminated FFI cut for aperture photometry \\
EXTVER & 1 & int & extension version \\
TIMESYS & `TDB' & str & TESS Barycentric Dynamical Time \\
BUNIT & `e-/s' & str & flux unit \\
STAR\_X & 1.511527631847869 & float & star x position in cut\tablenotemark{a} \\
STAR\_Y & 1.963850491691666 & float & star y position in cut\tablenotemark{a} \\
COMMENT &  &  & hdul{[}0{]}.data{[}:,star\_y,star\_x{]}=lc \\
ORIGIN & `UCSB/TGLC' & str & institution responsible for creating this file \\
TELESCOP & `TESS' & str & telescope \\
INSTRUME & `TESS Photometer' & str & detector type \\
FILTER & `TESS' & str & the filter used for the observations \\
OBJECT & `Gaia DR3 5707485527450614656' & str & string version of Gaia DR3 ID \\
GAIADR3 = & 5707485527450614656 & int & integer version of Gaia DR3 ID \\
TICID & `218795833' & str & TESS Input Catalog ID \\
SECTOR & 7 & int & observation sector \\
CAMERA & 2 & int & camera No. \\
CCD & 3 & int & CCD No. \\
CUT\_X & 0 & int & FFI cut x index \\
CUT\_Y & 0 & int & FFI cut y index \\
CUTSIZE & 90 & int & FFI cut size \\
RADESYS & `ICRS' & str & reference frame of celestial coordinates \\
RA\_OBJ & 124.6067520456133 & float & {[}deg{]} right ascension, J2000 \\
DEC\_OBJ = &  -19.66278772837456 & float & {[}deg{]} declination, J2000 \\
TESSMAG = & 14.54195107475864 & float & TESS magnitude, fitted by Gaia DR3 bands\tablenotemark{b} \\
GAIA\_G & 15.67702007293701 & float & Gaia DR3 g band magnitude \\
GAIA\_BP = & 17.19266128540039 & float & Gaia DR3 bp band magnitude \\
GAIA\_RP = & 14.48194599151611 & float & Gaia DR3 rp band magnitude \\
RAWFLUX = & 147.7626953125 & float & median flux of raw FFI \\
CALIB & `TGLC' & str & pipeline used for image calibration
\enddata
\tablenotetext{a}{Pixel position of the star in the 5*5 cutout if save\_aper=True}
\tablenotetext{b}{Caculated with Equation \ref{eq1}.}
\end{deluxetable}

\startlongtable
\begin{deluxetable}{llcl}
\tablecaption{Secondary Headers\tablenotemark{a} \label{tab:2header}}
\tablewidth{0pt}
\tablehead{\colhead{Header Card} &\colhead{Default/Example Value} &\colhead{Data Type} & \colhead{Description} }
\startdata
TIMEREF & `SOLARSYSTEM' & str & barycentric correction applied to times \\
TASSIGN & `SPACECRAFT' & str & where time is assigned \\
BJDREFI & 2457000 & int & integer part of BJD reference date \\
BJDREFR & 0.0 & float & fraction of the day in BJD reference date \\
TIMEUNIT & `d' & str & time unit for TIME \\
TELAPS & 24.41693964503611 & float & {[}d{]} TSTOP-TSTART \\
TSTART & 1491.661149617617 & float & {[}d{]} observation start time in TBJD \\
TSTOP & 1516.078089262653 & float & {[}d{]} observation end time in TBJD \\
MJD\_BEG & 58491.16114961762 & float & {[}d{]} start time in barycentric MJD \\
MJD\_END & 58515.57808926265 & float & {[}d{]} end time in barycentric MJD \\
TIMEDEL & 0.02248336983889145 & float & {[}d{]} time resolution of data \\
XPTIME & 1800 & int & {[}s{]} exposure time \\
PSF\_ERR & 3.37816157065132 & float & {[}e-/s{]} PSF flux error \\
APER\_ERR & 1.880864044956725 & float & {[}e-/s{]} aperture flux error \\
CPSF\_ERR & 0.01337824161713607 & float & {[}e-/s{]} calibrated PSF flux error\tablenotemark{b} \\
CAPE\_ERR & 0.007316015100534172 & float & {[}e-/s{]} calibrated aperture flux error\tablenotemark{b} \\
NEAREDGE & False & bool & distance to edges of FFI \textless{}= 2\tablenotemark{c} \\
LOC\_BG & -292.0021735884076 & float & {[}e-/s{]} locally modified background \\
COMMENT &  & str & TRUE\_BG = hdul{[}1{]}.data{[}'background'{]} + LOC\_BG \\
WOTAN\_WL & 1 & int & wotan detrending window length \\
WOTAN\_MT & `biweight' & str & wotan detrending method
\enddata
\tablenotetext{a}{We omit light curve extension headers and duplicate rows from the Primary header. Light curve extensions are discussed separately in \ref{appendix:a3}. }
\tablenotetext{b}{As discussed at the end of Section \ref{sec:exoplanet}, the calibrated aperture flux has an almost halved uncertainty compared to the calibrated PSF flux for this light curve. }
\tablenotetext{c}{NEAREDGE indicates whether the star is 2 pixels or closer to the edge of the FFI. If True, the PSF light curves can not be fitted, and only the aperture light curves are available.}
\end{deluxetable}

\subsection{Light curve extensions} \label{appendix:a3}
The light curve is stored in the second HDU as a binary table. All columns are listed in Table \ref{tab:extensions}. The calibrated fluxes are ready for transit detections; the uncalibrated fluxes are best for variable star sciences. The PSF light curves usually provide better deblending, but the aperture light curves offer higher precision and a more consistent amplitude if the target is in a crowded field. The background fit shows the background variation and could indicate stray light from the Earth and the Moon. The cadence number is the cadence of the FFI. The TESS flag follows the FFI convention \footnote{\url{https://outerspace.stsci.edu/display/TESS/2.0+-+Data+Product+Overview}}. The TGLC flag has only the first bit monitoring the presence of stray light, which is achieved by marking cadences with backgrounds at least five standard deviations from the median background. 

\begin{deluxetable}{clcl}[h!]
\tablecaption{Light curve extensions \label{tab:extensions}}
\tablewidth{0pt}
\tablehead{\colhead{Column} &\colhead{Name} &\colhead{Data Type} & \colhead{Description} }
\startdata
1 & time & numpy.ndarray & Time (TBJD) \\
2 & psf\_flux & numpy.ndarray & PSF flux ($e^-$/ s) \\
3 & aperture\_flux & numpy.ndarray & Aperture flux ($e^-$/ s) \\
4 & cal\_psf\_flux & numpy.ndarray & Calibrated PSF flux (normalized and detrended) \\
5 & cal\_aper\_flux & numpy.ndarray & Calibrated aperture flux (normalized and detrended) \\
6 & background & numpy.ndarray & Fitted background value ($e^-$/ s) \\
7 & cadence\_num & numpy.ndarray & FFI cadence number \\
8 & TESS\_flags & numpy.ndarray & FFI quality flags (directly from FFI) \\
9 & TGLC\_flags & numpy.ndarray & TGLC flags
\enddata
\end{deluxetable}

\clearpage
\bibliographystyle{aasjournal}
\bibliography{refs.bib}

\end{document}